\documentclass[12pt,letterpaper]{article}
\usepackage[top=1.25in, bottom=1.25in, left=1.25in, right=1.25in]{geometry}
\usepackage[utf8]{inputenc}
\usepackage{amsmath}
\usepackage{amsfonts}
\usepackage{amssymb}
\usepackage{amsthm}
\usepackage{caption}
\usepackage{subcaption}
\usepackage{graphicx}
\usepackage{float}
\usepackage[hypertexnames=false]{hyperref}
\usepackage[shortlabels]{enumitem}
\usepackage{pgfplots}
\usepackage{tikz,pgf}
\usetikzlibrary{calc,patterns,fillbetween}
\usetikzlibrary{decorations.pathreplacing,calligraphy}
	\pgfplotsset{compat=1.6}
\hypersetup{
  colorlinks=true,
  urlcolor=blue,
  linkcolor=blue,
  citecolor=blue}
\usepackage{import}
\usepackage{verbatim}
\usepackage{natbib}
\usepackage{mathtools}
\usepackage{bm}
\usepackage{standalone}
\allowdisplaybreaks

\providecommand{\keywords}[1]{\noindent\textit{Keywords:} #1}

\usepackage{setspace}
\usepackage{mdframed}
\newcommand{\rl}{L}
\newcommand{\rh}{H}
\newcommand{\m}{M}

\DeclareMathOperator*{\argmax}{arg\,max}
\DeclareMathOperator*{\supp}{supp}

\usepackage{amsthm,thmtools}
\declaretheorem[name=Theorem]{theo}
\declaretheorem[name=Corollary]{cor}
\declaretheorem[name=Lemma]{lem}
\declaretheorem[name=Benchmark,style=definition]{bench}

\title{\bf Monopoly agenda control with privately informed voters}
\author{\large Kirill S. Evdokimov\thanks{Department of Politics, Princeton University. Email: \url{ksevdokimov@princeton.edu}. I thank H\"{u}lya Eraslan for continuing support, Mallesh Pai and Nina Bobkova for many valuable suggestions that helped to improve the paper, and Isabelle Perrigne for detailed comments on early drafts. I also thank Germ\'{a}n Gieczewski, Matias Iaryczower, Maxim Ivanov, Navin Kartik, Mikhail Panov, Tomas Sj\"{o}str\"{o}m, Joel Sobel, Konstantin Sonin, and Joel Watson for useful suggestions. An earlier version of this paper circulated under the title ``The Coase conjecture and agreement rules in policy bargaining.''}}

\date{ \today}

\begin{document}

\begin{titlepage}
\maketitle
\thispagestyle{empty}
\setcounter{page}{0}

\begin{abstract}
  An agenda-setter repeatedly proposes a spatial policy to voters until some proposal is accepted. Voters have distinct but correlated preferences and receive private signals about the common state. I investigate whether the agenda-setter retains the power to screen voters as players become perfectly patient and private signals become perfectly precise. I show that the extent of this power depends on the relative precision of private signals and the conflict of preferences among voters, confirming the crucial role of committee setting and single-peaked preferences. When the private signals have equal precision, the agenda-setter can achieve the full-information benchmark. When one voter receives an asymptotically more precise signal, the agenda-setter's power to screen depends on preference diversity. These results imply that the lack of commitment to a single proposal can benefit the agenda-setter. Surprisingly, an increase in the voting threshold can allow the agenda-setter to extract more surplus.

\end{abstract}

\end{titlepage}

\section{Introduction}

Many collective decision-making institutions involve an agenda-setter with the sole power to make policy proposals, a group of voters with a collective power to veto proposals, and the ability to revise proposals if the previous ones are rejected. Examples abound, including the nomination and confirmation of presidential appointees in the United States, the decision-making in committees and boards, and the proposal of and voting on public spending measures.

When the agenda-setter is uncertain about voters' preferences, she can learn from rejected proposals and use this information when making revised proposals. In light of the \cite{Coase1972} conjecture, it is not clear if there is an agenda-setting advantage in such a setting. Famously, the Coase conjecture claims that a monopolist selling a durable good on the market would not be able to charge a price above the cost of production because of the competition with the future self. The Coase conjecture holds in veto bargaining over price.\footnote{For example, see \cite*{FLT} and \cite*{GSW}. This result is sensitive to the specific assumptions, and several authors later showed that the Coase conjecture can be violated even when bargaining over price, including \cite{AuDe1989}, \cite{DL}, and \cite{fuchs2013bridging}.} However, the examples of decision-making situations above do not fit this description since the bargaining is not over price but over policy, and the power to approve and veto proposals is not individual but collective. 

In this paper, I study a model of policy-making in which an agenda-setter repeatedly makes policy proposals until voters approve one of them. I focus on situations in which the agenda-setter is uncertain about the preferences of voters and learns from the rejected proposals. Voters receive private signals about the common state of the world and may reveal private information to the agenda-setter and other voters through their voting behavior. Thus, the agenda-setter faces the problem of screening the private information of voters who face both pivotal and signaling incentives when deciding how to vote. I investigate whether the agenda-setter has the ability to screen voters as the bargaining frictions disappear, how the voting rule affects equilibrium outcomes, and whether the ability to make a revised proposal is valuable to the agenda-setter. 

Strategic tensions in this model arise from the differences in goals among players, the asymmetric information about the effects of policies, and impatience. The agenda-setter is risk-neutral and maximizes the implemented policy as in \cite{RoRo79}. Voters have quadratic preferences with distinct state-dependent ideal policies and want to increase the status-quo policy. Thus, all players agree on the direction of the desired policy change but disagree on its magnitude.\footnote{These assumptions can be relaxed without affecting the qualitative results. For example, players can have any (continuous) quasi-concave utility function that is single-peaked. The only restriction is that the agenda-setter's ideal policy is greater than the highest policy that gives at least the status-quo level of utility to at least one voter.\label{foot1}} Voters receive private signals about the common state of the world, which can be either high or low.\footnote{As discussed below, the relative precision of private signals has an effect on the relative ability of voters to influence the beliefs of the agenda-setter and other voters.} The agenda-setter is uninformed and must infer the information about the common state from the actions of voters.\footnote{Equivalently, one can assume that all players receive a public signal about the state, and only after that does each voter receive an additional private signal.} Each voter's ideal policy is greater in the high state compared to the low state. If the state was known, the agenda-setter could implement a larger policy in the high state than in the low state. Since the state is not known, voters with high private signals may benefit from mimicking voters with low private signals, limiting the ability of the agenda-setter to extract surplus from the decisive voter.\footnote{Voters are ordered by their ideal policies, and this ordering does not depend on the state. When the required voting threshold is $q$, the voter with the $q$-th highest ideal policy is called decisive.} Finally, screening voters by the agenda-setter and mimicking lower types by voters are costly because players prefer earlier agreements and discount future payoffs.

\paragraph{Results}

The main technical results are Theorems \ref{theorem-equal} and \ref{theorem-unequal}, which characterize the limits of the agenda-setter's highest expected payoff when voters receive signals of equal precision (Theorem \ref{theorem-equal}) and when one voter receives a more precise signal (Theorem \ref{theorem-unequal}).\footnote{Despite several restrictions on equilibrium strategies, the model exhibits a severe multiplicity of equilibrium outcomes. The main reason is the lack of coordination among voters and the dynamic nature of the game. Since future proposals depend on history, it is often possible to support multiple equilibrium profiles using implicit punishments for unilateral deviations. Throughout the paper, I focus on characterizing the tight upper bound for the agenda-setter's expected payoff.} These results, along with the reasoning behind them, have the following implications.

Can the agenda-setter effectively screen voters' types? Coase conjectured that the agenda-setting advantage disappears when players become perfectly patient. The intuition behind this conjecture, initially formulated for a durable-good monopolist but later shown to hold in bilateral bargaining over price by \cite*{FLT}, is that the veto player evaluates the current proposals not with respect to the status quo but with respect to the anticipated revisions. Consequently, the agenda-setter competes with the future self, and this competition completely washes away the agenda-setting advantage. The model considered here differs from a typical model studied in the literature on the Coase conjecture because it considers a committee of voters with single-peaked preferences. Voters dislike extreme proposals on both sides of their ideal policy. If the policy that targets low types is far from the ideal policy of high types, high types might not find it desirable to mimic low types. In contrast, mimicking low types is always desirable when bargaining over a distributive policy when players are sufficiently patient. The requirement that proposals are approved by voting also creates additional incentives for players since voters face a coordination problem. Moreover, voters must take into account the effect of their actions on the beliefs of the agenda-setter and other voters.

I show that the Coase conjecture is violated when the bargaining is over a spatial policy or when bargaining against a committee. Under some conditions, the agenda-setter's highest expected payoff can be achieved by the equilibrium path that coincides with the one arising in bilateral bargaining.\footnote{In particular, when $q=1$ or when voter $i<q$ receives a more precise signal. Recall that voters are ordered with respect to their ideal policies, with the voter $i$ having the $i$-th highest ideal policy in each state.} In other words, the proposed policies and the acceptance probabilities are the same as they would be if a single voter replaced the committee. In fact, the equilibrium path is analogous to the bilateral bargaining over price, with one major difference. Since voters have single-peaked preferences, the initial screening proposal converges to a higher policy than is acceptable to the voter in the low state but is still accepted with certainty by the voter in the high state. In this case, the Coase conjecture can be violated because of the single-peaked preferences, and the committee setting is not essential.

However, the committee setting can have a profound effect on the agenda-setter's ability to screen voters. When voters receive private signals of equal precision and the voting threshold requires at least two voters to accept the proposal, the agenda-setter can extract all surplus from the decisive voter (and thus achieve the complete information Benchmark \ref{bench:full}) as the signals become perfectly precise and the players become perfectly patient. To achieve this, we must find an equilibrium in which the initial proposal is accepted in the high state with certainty and sets the decisive voter in the high state to the status-quo level of utility (both in the limit of patience and precision). The crucial step involves making voters collectively accept a policy that at least some of them would prefer to be collectively rejected. In this case, the Coase conjecture is violated because the veto power of voters is collective, and the single-peaked preferences are not essential.

How does an increase in the voting threshold affect the implemented policies and the agenda-setter's expected payoff?\footnote{A variety of voting thresholds are used in practice. The majority rule is commonplace. Several supermajority requirements are included in the United States Constitution. The ``unanimous written consent'' procedure many corporate boards use requires unanimity to support an action.} There is a tradeoff between different voting thresholds from the agenda-setter's perspective. On the one hand, a smaller threshold allows the agenda-setter to target fewer voters more aligned with the agenda-setter. In comparison, a larger threshold forces the agenda-setter to secure approval from more voters. On the other hand, a larger threshold may reduce credible revised proposals, providing a form of commitment for the agenda-setter. When the agenda-setter becomes sufficiently pessimistic about the state being high, she makes a proposal that targets the decisive voter in the low state. Such a proposal decreases with the voting threshold. Since voters with single-peaked preferences dislike extreme policies, this may allow the agenda-setter to extract more surplus from the decisive voter in the high state. I show that this mechanism comes into play when one voter receives an asymptotically more precise signal and the ideal policies of voters are moderately diverse.

Is the ability to make revised proposals valuable to the agenda-setter? When the agenda-setter makes a take-it-or-leave-if offer -- a situation analyzed in Benchmark \ref{bench:take} -- she either makes a proposal that is acceptable to the decisive voter in both states or a proposal that is only acceptable in the high state and otherwise preserves the status quo. It becomes clear that the ability to make revised proposals is valuable when the agenda-setter can extract all surplus from the decisive voter or when the agenda-setter can extract some surplus from the decisive voter in the high state and the prior belief that the state is high is sufficiently weak. In contrast, this ability is harmful when the agenda-setter cannot extract all surplus and the belief that the state is high is sufficiently strong.

\paragraph{Contributions}

This paper investigates the connections between single-peaked preferences, collective veto power, and the Coase conjecture.\footnote{The earliest formal work on the Coase conjecture includes \cite{Stokey1981} and \cite{Bulow1982}. More recent papers are (among others): \cite{DL}, who assume that the seller is privately informed about her cost; \cite{Or2017}, who allows the seller's cost to evolve stochastically; and \cite{DoSk2020}, who follow a mechanism design approach. In a setting with spatial policies, \citet*{Kartik} derive conditions for interval delegation to be an optimal mechanism without transfers from the agenda-setter's perspective.} In a bilateral setting, \cite*{ali2023sequential} show that single-peaked preferences allow equilibria without skimming, i.e., in which the policy proposals can increase on the path with non-trivial probability. In contrast, this paper focuses on equilibria with skimming (appropriately defined for committee setting and private signals) and emphasizes the role of voting on equilibrium outcomes.

Committee voting introduces novel challenges in assessing the agenda-setting power in the presence of asymmetric information and repeated proposals. The main results demonstrate that the relative precision of private signals has a profound effect on the agenda-setter's ability to screen voters. This paper also provides novel sufficient conditions for the efficient aggregation of private information in a finite election. Since the number of voters is finite and fixed, the full information equivalence relies on the assumption that the private signals become perfectly precise and voting strategies are monotonic.\footnote{After the ground-breaking paper by \cite{AuSm96}, the literature on information aggregation has mostly shifted to large elections. Information aggregation in large elections with two alternatives has been studied by \citet*{FePe97,My2000,persico2004committee,bhattacharya2013preference,Mc13,ekmekci2020manipulated,barelli2022full,kosterina2023information}, among others. An alternative approach to studying information aggregation in a finite election is to assume that the election is conducted after a round of non-binding voting or a communication stage (see \citet*{austen2009information} for an early survey). In this paper, communication is not allowed, and every round of voting in the dynamic model is binding.} Further, this paper introduces a novel approach to the analysis of the tradeoff between the signaling and pivotal incentives in a setting with a finite number of voters with asymptotically precise signals.\footnote{This tradeoff has been previously studied in a variety of contexts,\footnote{See \citet*{Pi2000}, \citet*{Ra03}, \cite{Me2005}, \cite{Sh2006}, \cite{MeTu2007}, \citet*{MeSh2009}, and \citet*{Mc17}, among others.} typically involving large elections with a fixed precision of private signals.}

Finally, this paper shows that in the absence of formal commitment power, restrictive decision-making rules that never benefit the agenda-setter in a static version of the model sometimes provide a form of commitment that helps the agenda-setter by influencing incentive constraints in future periods.\footnote{There is a larger literature that studies the role of the voting rule on equilibrium outcomes in collective decision-making. A partial list of papers with ``surprising'' results regarging the voting rule includes \cite{feddersen1998convicting}, \cite{chen2014rhetoric}, \cite*{diermeier2017political}, and \cite{eraslan2017some}.}

\section{Main ideas}\label{sec-ideas}

The main results of this paper rely on single-peaked preferences of voters, the collective power of voters to accept and reject policy proposals, or both. I assume that voters rank policies non-monotonically: each voter's payoff increases as the policy moves from the status quo level to the voter's ideal policy and then decreases when the policy moves further away from it. This lack of monotonicity is a crucial difference compared to models of trade, in which buyers always prefer a smaller price.

\subsection{Skimming equilibria with single-peaked preferences}\label{ideas:single}

For simplicity, suppose that there is a single voter $q$. As shown in the proof of Theorem \ref{theorem-unequal}, the characterization of the equilibrium path in this case coincides with the characterization of the equilibrium path when there are multiple voters in some special cases.\footnote{Namely, when $q=1$ or when voter $q$ receives a more precise private signal compared to other voters. The notion of relative signal precision is formalized in Section \ref{sec-model}.} Moreover, suppose that voter $q$ rates policies $p\in[0,M]$ using a state-dependent quadratic loss utility $u_q^\omega(p)=-(\frac12y^\omega_q-p)^2$ such that the ideal policies in states $\omega=\ell$ and $\omega=h$ satisfy $\frac12y^\ell_q<y^\ell_q<\frac12y^h_q$. The last inequality guarantees that voter $q$ in state $h$ prefers some policies greater than $y^\ell_q$ to $y^\ell_q$ itself. I refer to $y^\omega_q$ as the reservation policy of voter $q$ in state $\omega$. This is the largest policy that such a voter weakly prefers to the status quo.

When the voter has single-peaked preferences, there generally exist equilibria without skimming. In particular, an equilibrium path may exhibit leapfrogging, when early offers target the low type of voter and the revised proposals increase as the game progresses. \cite*{ali2023sequential} show that the expected implemented policy may exceed the lowest possible reservation policy in such equilibria.\footnote{More precisely, they show that in equilibria with leapfrogging, the agenda-setter can approximate the commitment expected payoff, where commitment means that the agenda-setter can commit to a sequence of proposals. Moreover, they show that leapfrogging is in some cases necessary to achieve this benchmark.} In this paper, I focus on equilibria with skimming, so the question remains what the agenda-setter can achieve in equilibria with a decreasing sequence of offers.\footnote{When there are multiple voters, the notion of skimming must be adjusted to allow an increase in the belief that the state is high and an increase in the offer, but with vanishing probability.} The answer is yes. The description of an equilibrium path is complicated by the fact that the voter does not directly observe the state. To simplify exposition, I describe the limit of an equilibrium path as signals become perfectly precise. The full construction of equilibrium strategies is given in the proof of Theorem \ref{theorem-unequal}.

For each prior belief $\mu^0(h)$, there is a (necessarily finite) sequence of policies $\{p_t\}_{t=1}^T$ and belief cutoffs $\{\m_t\}_{t=1}^T$ such that the agenda-setter follows a decreasing sequence of proposals $p_T>\dots>p_1=y^\ell_q$. In state $\ell$, every proposal except $p_1$ is rejected by the voter, and in state $h$, the voter accepts proposal $p_t$ with probability $\varphi_t$ such that $\varphi_1=1$ and $\varphi_t\in(0,1)$ for each $t=2,\dots,T$. If the current proposal $p_t$ is rejected, the agenda-setter revises the belief $\mu(h)$ using the Bayes rule to $\m_{t-1}$ and makes a revised proposal $p_{t-1}$. Clearly, when the state is $\ell$ and players become perfectly patient, the agenda-setter's payoff converges to $y^\ell_q$.

\begin{figure}[ht!]
  \begin{subfigure}[t]{\linewidth}\centering
  \begin{tikzpicture}[scale=1]

\draw[line width=0pt,color=white] (0,0) -- (15,0);

\draw[line width=.5pt] (0,0) -- (0,1/8) node[above] {$0$};
\draw[line width=.5pt] (6,0) -- (6,1/8) node[above] {$y^\ell_q$};

\draw[line width=1pt, orange] (9,-1.25) -- (10,-1.25) node[right, black] {$u_q^h(x_1)$};
\draw[line width=1pt,dashdotted] (9,-2) -- (10,-2) node[right] {$(1-\delta)u_q^h(0)+\delta u_q^h(x_2)$};

\draw[line width=.5pt,->] (0,0) -- (10.25,0) node[right] {$x_1,x_2$};
\draw[line width=.5pt,dashed] (5,0) -- (5,1/8) node[above] {$\frac12y^h_q$};
\draw[line width=1pt, domain=2:8, samples=100, orange] plot (\x,{-(5-\x)^2/5}) node[right] {};
\draw[line width=.5pt,dashed] (10,0) -- (10,1/8) node[above] {$y^h_q$};

\draw[line width=1pt, dashdotted, domain=2:8, samples=100] plot (\x,{-(1/10)*5^2/5-(9/10)*(5-\x)^2/5}) node[right] {};

\draw[line width=.5pt,dashed,color=black] (6,0) -- (6,-3.4/5);
\draw[line width=.5pt,dashed,color=black] (6,-3.4/5) -- (6.844,-3.4/5);
\draw[line width=.5pt,dashed,color=black] (6.844,-3.4/5) -- (6.844,0);
\draw[line width=.5pt,dashed,color=black] (6.844,0) -- (6.844,1/8) node[above] {$p_2$};

\end{tikzpicture}
  \caption{Case $\frac12y_q^h\leq y_q^\ell$. Policy $p_2$ converges to $y_q^\ell$ as $\delta\to1$, implying that the agenda-setter loses the ability to screen different types of voter $q$.}
  \label{util1}
  \vspace{.3in}
  \end{subfigure}\\
  \begin{subfigure}[t]{\linewidth}\centering
  \begin{tikzpicture}[scale=1]

\draw[line width=0pt,color=white] (0,0) -- (15,0);

\draw[line width=.5pt] (0,0) -- (0,1/8) node[above] {$0$};

\draw[line width=1pt, orange] (9,-1.25) -- (10,-1.25) node[right, black] {$u_q^h(x_1)$};
\draw[line width=1pt,dashdotted] (9,-2) -- (10,-2) node[right] {$(1-\delta)u_q^h(0)+\delta u_q^h(x_2)$};

\draw[line width=.5pt] (3,0) -- (3,1/8) node[above] {$y^\ell_q$};

\draw[line width=.5pt,->] (0,0) -- (10.25,0) node[right] {$x_1,x_2$};
\draw[line width=.5pt,dashed] (5,0) -- (5,1/8) node[above] {$\frac12y^h_q$};
\draw[line width=1pt, domain=2:8, samples=100, orange] plot (\x,{-(5-\x)^2/5}) node[right] {};
\draw[line width=.5pt,dashed] (10,0) -- (10,1/8) node[above] {$y^h_q$};

\draw[line width=1pt, dashdotted, domain=2:8, samples=100] plot (\x,{-(1/10)*5^2/5-(9/10)*(5-\x)^2/5}) node[right] {};

\draw[line width=.5pt,dashed,color=black] (3,0) -- (3,-5.8/5);
\draw[line width=.5pt,dashed,color=black] (3,-5.8/5) -- (7.408,-5.8/5);
\draw[line width=.5pt,dashed,color=black] (7.408,-5.8/5) -- (7.408,0);
\draw[line width=.5pt,dashed] (7.408,0) -- (7.408,1/8) node[above] {$p_2$};

\end{tikzpicture}
  \caption{Case $\frac12y_q^h>y_q^\ell$. Policy $p_2$ converges to $(y_q^h-y_q^\ell)>y_q^\ell$ as $\delta\to1$, implying that the agenda-setter retains the ability to screen different types of voter $q$.}
  \label{util2}
  \vspace{.3in}
  \end{subfigure}\\
  \begin{subfigure}[t]{\linewidth}\centering
  \begin{tikzpicture}[scale=1]

\draw[line width=0pt,color=white] (0,0) -- (15,0);

\draw[line width=.5pt] (0,0) -- (0,1/8) node[above] {$0$};

\draw[line width=1pt, orange] (9,-1.25) -- (10,-1.25) node[right, black] {$u_1^h(x_1)$};

\draw[line width=.5pt] (2.5,0) -- (2.5,1/8) node[above] {$y^\ell_{q+1}$};
\draw[line width=.5pt] (3.5,0) -- (3.5,1/8) node[above] {$y^\ell_{q}$};

\draw[line width=.5pt,->] (0,0) -- (10.25,0) node[right] {$x_1$};
\draw[line width=.5pt,dashed] (5,0) -- (5,1/8) node[above] {$\frac12y^h_1$};
\draw[line width=1pt, domain=2:8, samples=200, orange] plot (\x,{-(5-\x)^2/5}) node[right] {};
\draw[line width=.5pt,dashed] (8.5,0) -- (8.5,1/8) node[above] {$y^h_{q+1}$};
\draw[line width=.5pt,dashed] (10,0) -- (10,1/8) node[above] {$y^h_1$};


\draw[line width=.5pt,dashed,color=black] (2.5,-6.25/5) -- (2.5,0);
\draw[line width=.5pt,dashed,color=black] (3.5,-2.25/5) -- (3.5,0);

\draw[line width=.5pt,dashed,color=black] (7.5,-6.25/5) -- (7.5,0);
\draw[line width=.5pt,dashed,color=black] (6.5,-2.25/5) -- (6.5,0);

\draw[line width=.5pt,dashed,color=black] (2.5,-6.25/5) -- (7.5,-6.25/5);
\draw[line width=.5pt,dashed,color=black] (3.5,-2.25/5) -- (6.5,-2.25/5);

\draw[line width=.5pt,dashed] (7.5,0) -- (7.5,1/8) node[above] {$p_2^{q+1}$};
\draw[line width=.5pt,dashed] (6.5,0) -- (6.5,1/8) node[above] {$p_2^{q}$};

\end{tikzpicture}
  \caption{Voter $i\leq q$ has an asymptotically more precise signal. In this example, an increase in the required quota from $q$ to $(q+1)$ increases the limit policy implemented in state $h$ from $p_2^q=(y^h_i-y^\ell_q)$ to $p_2^{q+1}=(y^h_i-y^\ell_{q+1})$.}
  \label{util3}
  \end{subfigure}
  \caption{The determination of policy $p_2$ and its limit as $\tau\to1$ and  $\delta\to1$.}\label{util}
\end{figure}
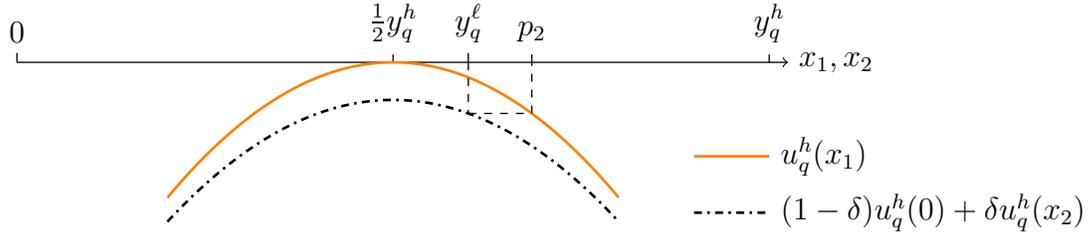
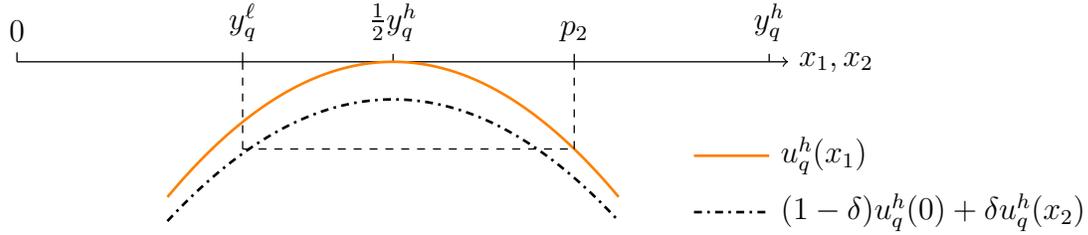
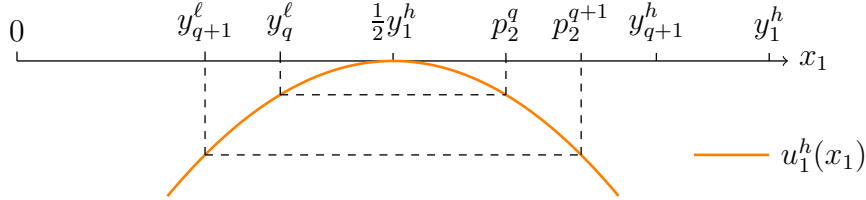

The voter in state $h$ is indifferent between accepting $p_t$ today and $p_{t-1}$ after a single period of delay; in particular, such voter is indifferent between accepting $p_2$ today and $y^\ell_q$ tomorrow. More precisely, policy $p_2$ is defined as $\max\{p\in[0,M]\mid u_q^h(p)\geq (1-\delta)u_q^h(0)+\delta u_q^h(y^\ell_q)\}$. By assumption, $y^\ell_q<\frac12y^h_q$ holds, and policy $p_2$ converges to $(y^h_q-y^\ell_q)$ as players become perfectly patient. In fact, $p_t$ converges $(y^h_q-y^\ell_q)$ and the probability of acceptance in state $h$ converges to 1 for each $t=2,\dots,T$. This implies that the agenda-setter's payoff in high state equals $(y^h_q-y^\ell_q)$, which is greater than $y^\ell_q$. In other words, the agenda-setter has the ability to screen voter types when the bargaining is over a policy and the voter's ideal policies in different states are sufficiently far apart (Figure \ref{util1}).

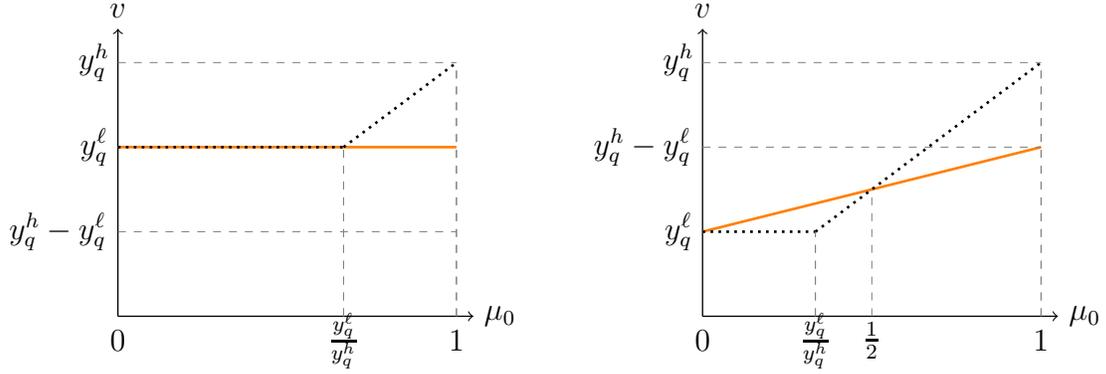
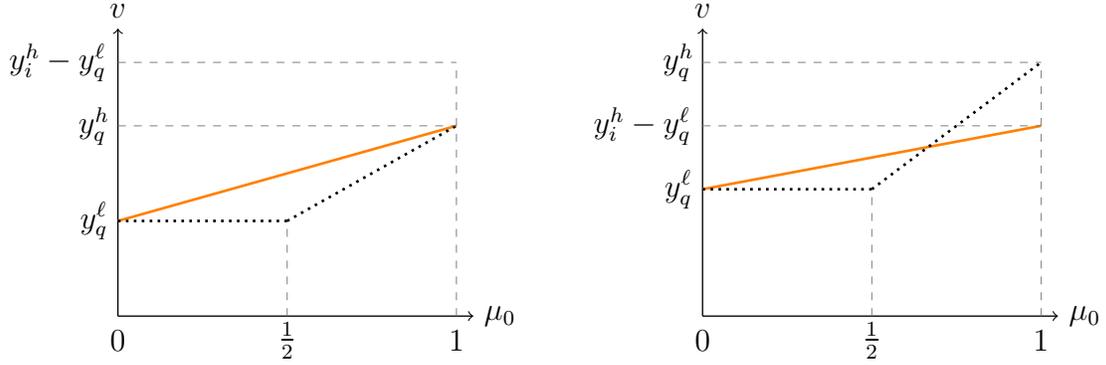
\begin{figure}[ht!]
  \begin{subfigure}[t]{.49\linewidth}
    \begin{tikzpicture}[scale=2.25]

\draw[line width=.5pt,->] (0,0) -- (2.1,0) node[right] {$\mu_0$};
\draw[line width=.5pt,->] (0,0) -- (0,1.7) node[above] {$v$};

\node[draw=none,fill=none] at (-1/7,1) {$y^\ell_{q}$};
\node[draw=none,fill=none] at (-1/7,1.5) {$y^h_{q}$};

\node[draw=none,fill=none] at (0,-1/7) {$0$};
\node[draw=none,fill=none] at (4/3,-1/7) {$\frac{y^\ell_{q}}{y^h_{q}}$};
\node[draw=none,fill=none] at (-5/14,0.5) {$y^h_{q}-y^\ell_{q}$};
\node[draw=none,fill=none] at (2,-1/7) {$1$};
\draw[dashed,line width=.25pt,gray] (4/3,0) -- (4/3,1);
\draw[dashed,line width=.25pt,gray] (2,0) -- (2,1.5);

\draw[line width=1pt,orange] (0,1) -- (2,1);


\draw[line width=1pt,dotted,black] (0,1) -- (4/3,1);
\draw[line width=1pt,dotted,black] (4/3,1) -- (2,1.5);

\draw[dashed,line width=.25pt,gray] (2/3*2,0) -- (2/3*2,1);
\draw[dashed,line width=.25pt,gray] (2,0) -- (2,1.5);
\draw[dashed,line width=.25pt,gray] (0,1.5) -- (2,1.5);
\draw[dashed,line width=.25pt,gray] (0,0.5) -- (2,0.5);



\end{tikzpicture}
    \caption{Voter $q$ has a more precise signal and $y^h_q\leq2y^\ell_q$; or voter $i<q$ has a more precise signal and $y^h_i-y^h_q\leq y^\ell_q$.
    }\label{fig-coasian-a}
  \end{subfigure}\hfill
  \begin{subfigure}[t]{.49\linewidth}
    \begin{tikzpicture}[scale=2.25]

\draw[line width=.5pt,->] (0,0) -- (2.1,0) node[right] {$\mu_0$};
\draw[line width=.5pt,->] (0,0) -- (0,1.7) node[above] {$v$};

\node[draw=none,fill=none] at (-1/7,.5) {$y^\ell_{q}$};
\node[draw=none,fill=none] at (-1/7,1.5) {$y^h_{q}$};
\node[draw=none,fill=none] at (-5/14,1) {$y^h_{q}-y^\ell_{q}$};

\node[draw=none,fill=none] at (0,-1/7) {$0$};
\node[draw=none,fill=none] at (2/3,-1/7) {$\frac{y^\ell_{q}}{y^h_{q}}$};
\node[draw=none,fill=none] at (2,-1/7) {$1$};
\draw[dashed,line width=.25pt,gray] (2/3,0) -- (2/3,0.5);
\draw[dashed,line width=.25pt,gray] (2,0) -- (2,1.5);

\draw[line width=1pt,orange] (0,.5) -- (2,1);


\draw[line width=1pt,dotted,black] (0,.5) -- (2/3,.5);
\draw[line width=1pt,dotted,black] (2/3,.5) -- (2,1.5);

\draw[dashed,line width=.25pt,gray] (1/3*2,0) -- (1/3*2,.5);
\draw[dashed,line width=.25pt,gray] (2,0) -- (2,1.5);
\draw[dashed,line width=.25pt,gray] (0,1.5) -- (2,1.5);
\draw[dashed,line width=.25pt,gray] (0,1) -- (2,1);

\draw[dashed,line width=.25pt,gray] (1,0) -- (1,.75);
\node[draw=none,fill=none] at (1,-1/7) {$\frac12$};

\end{tikzpicture}
    \caption{Voter $q$ has a more precise signal and $y^h_q>2y^\ell_q$.
    }\label{fig-coasian-b}
  \end{subfigure}\\
  \begin{subfigure}[t]{.49\linewidth}
    \begin{tikzpicture}[scale=2.25]

\draw[line width=.5pt,->] (0,0) -- (2.1,0) node[right] {$\mu_0$};
\draw[line width=.5pt,->] (0,0) -- (0,1.7) node[above] {$v$};

\node[draw=none,fill=none] at (-5/14,1.5) {$y^h_{i}-y^\ell_{q}$};
\node[draw=none,fill=none] at (-1/7,0.5625) {$y^\ell_{q}$};
\node[draw=none,fill=none] at (-1/7,1.125) {$y^h_{q}$};

\node[draw=none,fill=none] at (0,-1/7) {$0$};
\node[draw=none,fill=none] at (2,-1/7) {$1$};

\draw[line width=1pt,orange] (0,0.5625) -- (2,1.125);

\draw[line width=1pt,dotted,black] (0,0.5625) -- (1,0.5625);
\draw[line width=1pt,dotted,black] (1,0.5625) -- (2,1.125);

\draw[dashed,line width=.25pt,gray] (0,1.5) -- (2,1.5);
\draw[dashed,line width=.25pt,gray] (0,1.125) -- (2,1.125);

\draw[dashed,line width=.25pt,gray] (2,0) -- (2,1.5);
\draw[dashed,line width=.25pt,gray] (1,0) -- (1,0.5625);

\node[draw=none,fill=none] at (1,-1/7) {$\frac12$};

\end{tikzpicture}
    \caption{Voter $i<q$ has a more precise signal and  $y^h_i-y^\ell_q\geq y^h_q$.
    }\label{fig-coasian-c}
  \end{subfigure}\hfill
  \begin{subfigure}[t]{.49\linewidth}
    \begin{tikzpicture}[scale=2.25]

\draw[line width=.5pt,->] (0,0) -- (2.1,0) node[right] {$\mu_0$};
\draw[line width=.5pt,->] (0,0) -- (0,1.7) node[above] {$v$};

\node[draw=none,fill=none] at (-1/7,1.5) {$y^h_{q}$};
\node[draw=none,fill=none] at (-1/7,0.75) {$y^\ell_{q}$};
\node[draw=none,fill=none] at (-5/14,1.125) {$y^h_{i}-y^\ell_{q}$};

\node[draw=none,fill=none] at (0,-1/7) {$0$};
\node[draw=none,fill=none] at (2,-1/7) {$1$};

\draw[line width=1pt,orange] (0,0.75) -- (2,1.125);

\draw[line width=1pt,dotted,black] (0,0.75) -- (1,0.75);
\draw[line width=1pt,dotted,black] (1,0.75) -- (2,1.5);

\draw[dashed,line width=.25pt,gray] (0,1.5) -- (2,1.5);
\draw[dashed,line width=.25pt,gray] (0,1.125) -- (2,1.125);

\draw[dashed,line width=.25pt,gray] (2,0) -- (2,1.5);
\draw[dashed,line width=.25pt,gray] (1,0) -- (1,0.75);

\node[draw=none,fill=none] at (1,-1/7) {$\frac12$};

\end{tikzpicture}
    \caption{Voter $i<q$ has a more precise signal and  $y^\ell_q<y^h_i-y^\ell_q<y^h_q$.
    }\label{fig-coasian-d}
  \end{subfigure}
  \caption{The agenda-setter's expected payoff when bargaining over policy. The dotted line represents the take-it-or-leave-it benchmark, and the solid line represents the agenda-setter's highest expected payoff, both in the limit as $\tau_j\to1$ for all $j\in\mathcal N$.}\label{fig-coasian}
\end{figure}

When the agenda-setter commits to a single proposal, the expected payoff equals $V^T_A=\max\{y^\ell_q,\mu^0(h) y^h_q\}$, where $T$ stands for a take-it-or-leave-it offer (see Benchmark \ref{bench:take}). It is interesting to compare this payoff with the agenda-setter's expected payoff when revisions are allowed (Figure \ref{fig-coasian}). In the model of trade, the ability to commit to a single proposal is beneficial for the agenda-setter, and the same is true when the ideal policies of voter $q$ in different states are not too far apart, $y^\ell_q\geq \frac12y^h_q$. This is a direct consequence of the forces behind the Coase conjecture. With commitment, the agenda-setter receives $y^h_q$ in the high state and $0$ in the low state when the prior belief $\mu(h)$ is sufficiently strong. Without commitment, the agenda-setter cannot screen the voter's types and receives $y^\ell_q$ in both states (Figure \ref{fig-coasian-a}).

In contrast, when $y^\ell_q<\frac12y^h_q$ and the prior belief $\mu(h)$ is sufficiently weak, the agenda-setter's expected payoff is strictly greater without commitment to a single proposal. In this case, the agenda-setter retains the ability to screen the voter's types and receives $(y^h_q-y^\ell_q)$ and $y^\ell_q$ in high and low states, respectively (Figure \ref{fig-coasian-b}).

\subsection{Exploiting the collective power of voters}\label{ideas:similarity}

The collective power of voters to accept and reject proposals is another essential feature of the model. I assume that the agenda-setter and voters observe the voting record and use this information to update their beliefs. Since voters receive private signals, the agenda-setter infers information from the voting behavior and uses it to update the belief about the state of the world. Therefore, voters are incentivized to use their actions to influence the belief of the agenda-setter. Moreover, voters also influence the beliefs of other voters and obtain equilibrium information about the state by conditioning on the others' voting behavior.\footnote{The importance of equilibrium information in voting was first noticed by \cite{Au1990}.} Compared to models of bilateral bargaining, proposals can be rejected by different coalitions of voters, leading to different updated beliefs. Despite various motives affecting voters, it is possible to disentangle these incentives and show that the agenda-setter can exploit the conflict of preferences among voters and their lack of coordination. The agenda-setter's ability to extract surplus from voters depends on the relative precision of private signals. When the signals have equal precision, the agenda-setter can exploit the lack of coordination among voters to achieve the complete information benchmark. This claim is a major component of Theorem \ref{theorem-equal} and does not rely on the single-peaked preferences of voters.

To illustrate, suppose that $q\geq2$, meaning that at least two voters must vote in favor of a policy for it to be implemented. Consider a tentative equilibrium in which the agenda-setter with belief $\mu$ proposes policy $\tilde p_q$, which will be defined shortly and generally depends on the prior belief $\mu$ and the precision of private signals $\tau$. Suppose that voters in $\{1,\dots,q-2\}$ accept and voters in $\{q+1,\dots,N\}$ reject this policy irrespective of their signals. At the same time, voters $(q-1)$ and $q$ only accept $\tilde p_q$ after receiving a signal that the state is $h$ and reject it after receiving a signal that the state is $\ell$. I will show that these voting strategies are sequentially rational and that policy $\tilde p_q$ converges to $y^h_q$, implying that the agenda-setter's expected payoff converges to $\mu(h) y^h_q+\mu(\ell)y^\ell_q$.

The agenda-setter's belief following a rejection of $\tilde p_q$ depends on the voting record. If both voters in $\{q-1,q\}$ reject $\tilde p_q$ (along with voters in $\{q+1,\dots,N\}$), the public belief $\mu$ is revised towards putting less weight on state $h$. When the signal precision is sufficiently high, the revised belief $\tilde \mu$ is close to zero by the Bayes rule, and therefore, the agenda-setter makes a revised proposal $p_q$, which converges to $y^\ell_q$ and is accepted in both states with certainty.\footnote{For the clarity of exposition, I assume that $p_q$ is accepted with probability 1. As shown in Lemma \ref{lem:screening}, this probability must converge to 1. I define $\tilde p_q$ for the general case in the proof of Theorem \ref{theorem-equal}.} If exactly one voter in $\{q-1,q\}$ accepts $\tilde p_q$, the revealed signals cancel each other, and the public belief $\mu$ remains the same.

Policy $\tilde p_q$ is defined as the largest policy $p$ that satisfies the following inequalities for both $i\in\{q-1,q\}$:
\begin{align}
  0 \geq\; & \mu(\ell)\tau^2\big[\delta V_{i}^{\ell,\mu}-\delta u^\ell_{i}(p_q)\big] + \mu(\ell)\tau(1-\tau)\big[u_{i}^\ell(p)-(1-\delta)u^\ell_{i}(0)-\delta V_{i}^{\ell,\mu}\big] \nonumber\\
  +\; & \mu(h)(1-\tau)^2\big[\delta V_{i}^{h,\mu} - \delta u^h_{i}(p_q)\big] + \mu(h)(1-\tau)\tau\big[u^h_{i}(p) - (1-\delta)u^h_{i}(0) - \delta V_{i}^{h,\mu}\big],  \label{def:tilde-p-2} \\
  0 \geq \; & \mu(\ell)(1-\tau)^2\big[\delta u^\ell_{i}(p_q) - \delta V_{i}^{\ell,\mu}\big] + \mu(\ell)(1-\tau)\tau\big[(1-\delta)u^\ell_{i}(0)+\delta V_{i}^{\ell,\mu} - u_{i}^\ell(p)\big] \nonumber\\
  +\; & \mu(h)\tau(1-\tau)\big[\delta u^h_{i}( p_q) - \delta V_{i}^{h,\mu}\big] + \mu(h)\tau^2\big[(1-\delta)u^h_{i}(0)+\delta V_{i}^{h,\mu} - u^h_{i}(p)\big], \label{def:tilde-p-1}
\end{align}
where $V_{i}^{\omega,\mu}$ is the expected payoff of voter $i$ in state $\omega$ if the game moves to the next period with the same public belief $\mu$ so that $\tilde p_q$ is proposed again. The right-hand sides of \eqref{def:tilde-p-2} and \eqref{def:tilde-p-1} are the gains in the expected payoff of voter $i$ with signals $s_i=\rl$ and $s_i=\rh$ from deviating. Using the continuation strategies of voters, we can find $V_{i}^{\ell,\mu}=\frac{1}{1-2(1-\tau)\tau\delta}\left[(1-\tau)^2 u_{i}^\ell(\tilde p_q) + \tau^2\delta u_{i}^\ell(p_q) + \tau(2-\tau)(1-\delta)u^\ell_{i}(0)\right]$, which converges to $(1-\delta)u^\ell_i(0)+\delta u^\ell_i(y^\ell_q)$ as the signals become perfectly precise. We can also find $V_{i}^{h,\mu}=\frac{1}{1-2(1-\tau)\tau\delta}\left[\tau^2 u_{i}^h(\tilde p_q) + (1-\tau)^2\delta u_{i}^h(p_q) + (1-\tau^2)(1-\delta)u^h_{i}(0)\right]$, which converges to $u_{i}^h(\lim_{\tau\to1}\tilde p_q)$.

By construction, voters in $\{q-1,q\}$ do not have the incentive to deviate from accepting $\tilde p_q$ after receiving high signals and rejecting $\tilde p_q$ after receiving low signals.\footnote{It is shown that other voters have no profitable deviations in the proof of Theorem \ref{theorem-equal}.} To prove that $\tilde p_q$ must converge to $y^h_q$ as the signals become perfectly precise, take any sequence $\{\tau_n\}$ converging to 1 and suppose inequality \eqref{def:tilde-p-1} binds infinitely often along the sequence. As $\tau_n$ converges to 1, the first three components on the right-hand side of \eqref{def:tilde-p-1} converge to 0, and therefore the fourth component must also converge to 0. Pluggning in the limit of $V_q^{h,\mu}$ derived above, we obtain $\lim_{\tau\to1}\tilde p_q=y^h_q$. In the complementary case, there exists $m$ such that \eqref{def:tilde-p-2} binds for all $n\geq m$. However, by the same argument \eqref{def:tilde-p-1} cannot hold if the limit of $\tilde p_q$ is below $y^h_q$.

Notice that both voters $(q-1)$ and $q$ may prefer a policy close to $y^\ell_q$ to a policy close to $y^h_q$ when they believe that the state is $h$, and therefore would like to jointly reject $\tilde p_q$ after receiving high signals in order to convince the agenda-setter that the state is $\ell$. However, if one voter receives a signal that the state is $h$, she believes that the other voter has also received a signal that the state is $h$ and therefore expects the other voter to vote in favor of $\tilde p_q$. A unilateral deviation to reject $\tilde p_q$ leads to a delay and a continuation play in which $\tilde p_q$ is implemented almost certainly, making such deviation not profitable.

\subsection{Exploiting the conflict in preferences}\label{ideas:conflict}

When one voter receives an asymptotically more precise private signal, the agenda-setter's ability to screen voters depends on the conflict in preferences among voters. An asymptotically more precise signal is defined in Section \ref{sec-model} and requires that the precision of voter $i$'s signal converges to 1 sufficiently fast relative to the precisions of other voters. Full extraction of surplus from the decisive voter is still possible but requires a different argument. The argument in the previous section relies on the assumption that private signals have identical precision so that two contradicting signals of voters $(q-1)$ and $q$ cancel each other after being revealed by truthful voting strategies.\footnote{This assumption can be relaxed, but in order to achieve a complete information benchmark, this assumption must hold asymptotically as signals become perfectly precise.} When voter $i\in\{q-1,q\}$ receives a more precise private signal, then following conflicting votes by $(q-1)$ and $q$, the revised proposal is close to $y^\ell_q$ (instead of $y^h_q$ as shown in the previous section). Therefore, voter $i$ with a high signal has a profitable deviation when she prefers $y^\ell_q$ to policy $y^h_q$ in state $h$, which may or may not be the case depending on the preferences of voters.

More generally, when voter $i\leq q$ receives a more precise signal than other voters, the highest policy that the agenda-setter can propose that is accepted in state $h$ converges to $\min\{y^h_q,y^h_i-y^\ell_q\}$, as shown in the proof of Theorem \ref{theorem-unequal}. Therefore, the agenda-setter can achieve the complete information benchmark only when voter $i$ in state $h$ prefers policy $y^h_q$ to $y^\ell_q$ (Figure \ref{fig-coasian-c}). Notice that this condition cannot be satisfied when $i=q$, and is violated when voters $i$ and $q$ have similar preferences in state $h$ (Figure \ref{fig-coasian-d}). Therefore, the agenda-setter benefits from the conflict in voters' preferences when one of them has a more precise signal.

\section{The model}\label{sec-model}

The agenda-setter $A$ and $N$ voters indexed by $i\in\mathcal N=\{1,\dots,N\}$ bargain over a one-dimensional policy $x\in[0,M]$ using a $q$-majority voting rule with $q\in\mathcal N$. The status-quo policy is $0$. For example, policy $x$ may represent a level of public spending or an increase in the capital stock of a company. In every period $t\in\{1,2,\dots,\infty\}$ some policy $x_t\in[0,M]$ is implemented. Period-$t$ policy $x_t$ is the status-quo policy 0 in every period until another policy $p\in[0,M]$ receives at least $q$ votes and gets implemented in all remaining periods.

After the agenda-setter makes a policy proposal $p_t$, voters observe $p_1$ and simultaneously cast their votes. The action set of each voter $i$ is $A_i=\{0,1\}$ with a generic element $a_{i,1}$, where $a_{i,1}=0$ if voter $i$ rejects the period-$t$ proposal and $a_{i,1}=1$ if voter $i$ accepts it. Therefore, voters cannot abstain. Proposal $p_t$ passes when the required quota $q$ of voters accepts it. In case $p_t$ passes, it is implemented in every period $s=t,t+1,\dots,\infty$, and the game ends. In case $p_t$ is rejected, the status quo is implemented in period $t$, $x_t=0$, and the same process is repeated in period $(t+1)$ after the voting record is revealed to the agenda-setter and all voters.

The agenda-setter is maximizing the policy -- the period utility of the agenda-setter from implementing policy $x\in[0,M]$ is $u_A(x)=x$. Voters have quadratic preferences over $[0,M]$ with the ideal policies that depend on state $\omega\in\Omega$. The period utility of voter $i\in N$ from implementing policy $x\in [0,M]$ in state $\omega\in\Omega$ is $u_i(x;\omega)=-\left(\frac12 y_i^{\omega}-x\right)^2$, where $\frac12 y_i^{\omega}>0$ is the ideal policy of voter $i$ in state $\omega$. Future payoffs are discounted at rate $\delta\in(0,1)$, so the total payoffs of the agenda-setter and voter $i\in N$ from a policy sequence $\bm x=\{x_t\}_{t=1}^\infty$ in state $\omega\in\Omega$ are $U_A(\bm x)=(1-\delta)\sum_{t=1}^\infty\delta^{t-1} u_A(x_t)$ and $U_i(\bm x;\omega)=(1-\delta)\sum_{t=1}^\infty\delta^{t-1} u_i(x_t;\omega)$.

\paragraph{States and signals}
There are two possible states, $\Omega=\{\ell,h\}$. I assume that every voter has monotone ideal policies in the sense that for each $q\in\mathcal N$ we have $0<y_q^\ell<y_q^h$. Therefore, state $\ell$ can be interpreted as being low and state $h$ as being high.\footnote{Note that if the state was known, the agenda-setter would be able to implement policy $y^\omega_q$ in state $\omega$ and therefore would prefer state $h$ over state $\ell$. See Benchmark \ref{bench:full}.}

Voters observe private signals about the state, but the agenda-setter does not. Each voter receives a binary signal $s_i\in S=\{\rl,\rh\}$. The signals are independent and identically distributed conditional on state $\omega$. The probability that voter $i\in N$ receives a signal $s_i=\rl$ in state $\ell$ and a signal $s_i=\rh$ in state $h$ are both $\tau_i\in(\frac12,1)$. In other words, $\tau_i\in(\frac12,1)$ can be interpreted as the precision of an individual signal $s_i$. The state is drawn and the private signals are observed at the beginning of the game.

For future reference, say that voter $i\in\mathcal N$ receives an \textit{asymptotically more precise} signal than other voters if given $N$ sequences of precision parameters $\{\tau^n_j\}_{n=1}^\infty$, $j\in\mathcal N$, each converging to 1, we have $\lim_{n\to\infty}\frac{1-\tau_i^n}{\prod_{j\neq i}(1-\tau_j^n)}=0$.

\paragraph{Strategies and beliefs}

The information sets of players lump together the histories that the players cannot distinguish, namely, those that only differ in the state and the signals of other players.\footnote{I carefully define histories and stationary Markov strategies in Appendix \ref{appendix-strategies}.} Therefore, players must form beliefs about the state and the signals of other players. To preserve tractability, I assume that in every period each voter $i\in\mathcal N$ is replaced by an identical voter with the same voting strategy but a newly drawn signal.\footnote{To emphasize, voters remain forward-looking and care about the complete policy sequence.} Then, players must form beliefs only about the state. Since the agenda-setter is uninformed and updates the belief using public information, the agenda-setter's belief is called \textit{public}. Notice that the public belief is the prior that voters update using their private signals.

A Markov proposal strategy $\pi$ depends on the history of proposals and votes only through the public belief about state $\omega$. Voter $i$'s Markov voting strategy $\alpha_i$ depends on the private signal $s_i$ and the public belief $\mu$. Therefore, Markov strategies of the agenda-setter and voter $i\in\mathcal N$ can be written as $\pi:\Delta(\Omega)\to\Delta([0,M])$ and $\alpha_i:S_i\times\Delta(\Omega)\times[0,M]\to[0,1]$. Notice that the Markov strategies defined here are also stationary since they do not depend on calendar time $t$.

\paragraph{Induced outcomes}

For each period $t$, public belief $\mu$, and state $\omega$, an assessment $(\sigma,\mathcal M)$ induces a (continuation) distribution $F^{\omega,\mu}$ over the space of policy sequences $[0,M]^\infty$. The expected payoffs of players for a given assessment $(\sigma,\mathcal M)$ can be written as $V_A(\mu)=\sum_{\omega\in\Omega}\mu(\omega)\int U_A(\bm x)\mathrm dF^{\omega,\mu}(\bm x)$ and $V_i(s_i,\mu)=\sum_{\omega\in\Omega}\mu(\omega\mid s_i)\int U_i(\bm x\mid \omega)\mathrm dF^{\omega,\mu}(\bm x)$ for $i\in\mathcal N$.

In some cases, it is more convenient to work directly with the distribution $G^{\omega,\mu}$ over pairs $(t,p)$, meaning that policy $p$ is accepted in period $t$. For each proposal $p\in[0,M]$, define $H(p)=\{(t,p')\mid t\in\{1,\dots,\infty\} \text{ and } p'>p\}$, which is the set of period-policy pairs $(t,p')$ where $p'$ is greater than $p$. Then, we can write a probability that a policy greater than $p$ is implemented at some time given the prior belief $\mu$ as $\sum_{\omega\in\Omega}\mu(\omega)G^{\omega,\mu}(H(p))$.

\paragraph{Equilibrium concept}

The equilibrium concept is a \textit{stationary Markov perfect Bayesian equilibrium} (an equilibrium). A stationary Markov strategy profile $\sigma$ is an equilibrium if there exists a belief system $\mathcal M$ such that $\sigma$ is sequentially rational given $\mathcal M$ and $\mathcal M$ is consistent with $\sigma$. In addition, I require that players ``cannot signal what they do not know,'' i.e., that the public belief does not directly depend on the proposed policies, and that the public belief is not updated after 0-probability events.

\paragraph{Coase conjecture}

The Coase conjecture states that as bargaining frictions disappear, the agenda-setter has no power to screen voters and equilibrium outcomes are efficient. Formally, the Coase conjecture is defined in terms of the agfenda-setter's expected payoff as follows:
\begin{equation}\label{coase-conjecture}
  \lim_{\delta\to1}\lim_{\tau\to1}V_A(\mu^0) = y^\ell_q.
\end{equation}
With a single voter $q$ who observes the state, this definition reduces to the familiar $\lim_{\delta\to1} V_A(\mu^0)=y^\ell_q$.

\paragraph{Skimming dynamics}

In a trade setting, the skimming dynamics involve a strictly decreasing sequence of proposals and a skimming property, which states that higher types purchase the good earlier. This definition cannot be directly applied to the setting with collective veto power and private signals about the state because of signal noise and the lack of coordination among voters.\footnote{If voters reveal their signals to each other and coordinate their actions, then the skimming property can be redefined in terms of probabilities of passing a given policy proposal in different states. Of course, this leaves the question of whether this is incentive-compatible for voters to reveal their signals and stick to the agreed-upon voting strategy.}

In this paper, the skimming property is replaced with several restrictions on voting strategies and equilibrium dynamics. Assume that voters use monotone voting strategies, so that for each voter $i\in\mathcal N$, public belief $\mu\in\Delta(\Omega)$, and proposal $p\in[0,M]$, we have:
\begin{equation}\label{coase-monotone}
  \alpha_i(p;\rh,\mu)\geq\alpha_i(p;\rl,\mu).
\end{equation}
Given monotone voting strategies, any additional vote in favor of a proposal increases the posterior belief that the state is high (Lemma \ref{belief:increasing}). Nonetheless, this posterior belief can still be above the prior even when the proposal fails to gather the required support, in contrast with bilateral setting.

Further, assume that for each $p\in[0,M]$, voting profile $a\in\{0,1\}^N$ such that $\sum_{i\in\mathcal N}a_i<q$, and public belief $\mu\in\Delta(\Omega)$, we have:
\begin{equation}\label{coase-dynamics}
  \lim_{\tau\to1}G^{\omega,\tilde\mu^{p,a}(\mu)}(H(p))=0,
\end{equation}
which implies that the on-path sequences of proposals are monotone with probability approaching one as signals become arbitrarily precise.

Finally, assume that taking two public beliefs $\mu$ and $\mu'$ such that $\mu(h)\leq \mu'(h)$, for each $t\in\{1,\dots,\infty\}$ and $p\in[0,M]$ we have:
\begin{equation}\label{coase-stochastic}
  \lim_{\tau\to1}G^{\omega,\mu}(H(p))\geq \lim_{\tau\to1}G^{\omega,\mu'}(H(p)).
\end{equation}
Together, these assumptions capture the essence of the skimming dynamics in the setting with collective veto power.

\subsection{Benchmarks}

\subsubsection{Complete information}\label{bench-full}

When the agenda-setter and voters observe the state, the agenda-setter can extract all surplus from the decisive voter. By a standard argument, in any stationary subgame-perfect equilibrium with weakly undominated voting strategies, each voter $i\in\mathcal N$ in state $\omega\in\Omega$ accepts proposal $p\in[0,M]$ if and only if $p\leq y^\omega_q$. In turn, the agenda-setter makes proposal $y^\omega_q$ with certainty. Given the voting strategies, this proposal is accepted without delay.

\begin{bench}[Complete information]\label{bench:full}
Fix the required quota $q$. When the information is complete, the agenda-setter's expected payoff equals $V^C_A(\mu^0)=\mu^0(h) y^h_q+\mu^0(\ell)y^\ell_q$.
\end{bench}

Lemma \ref{lem:bounds} in Appendix \ref{sec-proofs} shows that the complete information benchmark provides an upper bound for the agenda-setter's expected payoff in the baseline model with incomplete information and infinitely many periods. Moreover, in the next section, I show that the agenda-setter can approximate this upper bound under various conditions summarized in Corollary \ref{corollary}. It follows that this benchmark is the worst equilibrium outcome for voter $q$ because she receives her status-quo payoff in each state. Under informational asymmetry, voter $q$ receives an information rent in state $h$ in most equilibria. For instance, this happens in every equilibrium when the Coase conjecture holds; see Corollary \ref{corollary} for a precise statement of sufficient conditions.

\subsubsection{Take-it-or-leave-it offer}\label{bench-take}

If the agenda-setter can only make a single proposal, the status quo policy is implemented when the proposal is rejected. In this case, voters are not concerned with influencing future offers and only consider pivotal incentives given their signals and the strategies of other voters. Consider the difference in expected payoffs of voter $i$ from voting to accept and voting to reject proposal $p$ given signal $s_i$ and belief $\mu$:
\begin{align}
  \mathbb EU_i(p;s_i,\mu,\alpha_{-i}) \propto \sum_{\omega\in\{\ell,h\}}\mu(\omega)\tau_i^\omega(s_i) & \mathbb P_i(q-1\mid \omega)(u_i^\omega(p)-u_i^\omega(0)), \label{difference1}
\end{align}
where $\tau_i^\omega(s_i)$ is the probability that voter $i$ observes signal $s_i$ in state $\omega$,\footnote{For instance, $\tau_i^\ell(\rl)=\tau_i^h(\rh)=\tau_i$ and $\tau_i^\ell(\rh)=\tau_i^h(\rl)=1-\tau_i$.} and $\mathbb P_i(q-1\mid \omega)$ is the probability that player $i$ is pivotal in state $\omega$, i.e., $(q-1)$ other players vote to accept the proposal. As shown in Lemma \ref{benchmark:belowabove} in Appendix \ref{appendix-bench2}, from \eqref{difference1}, we can almost immediately conclude that proposals below $y^\ell_q$ are accepted by voters with $q$ highest ideal policies irrespective of their signals, and proposals above $y^h_q$ are rejected by voters with $(N-q+1)$ lowest ideal policies irrespective of their signals. Therefore, for each precision vector $\tau$, policies below $y^\ell_q$ are certainly accepted, and policies above $y^h_q$ are certainly rejected.

In contrast, the probability that a policy between $y^\ell_q$ and $y^h_q$ is accepted depends on the precision of private signals and does so for several reasons. The evaluation of policies by voters depends on their private beliefs about the state, which rely on the accuracy of private signals and influence the equilibrium strategies of voters, as captured by $\tau^\omega_i(s_i)$ in \eqref{difference1}. Moreover, when voters use informative strategies, the probability of receiving incorrect signals depends on $\tau$ as well. Such dependence on $\tau$ is eliminated by considering the limit of equilibria along a sequence of signal precisions converging to 1. Another complication is that voters receive equilibrium information. Since the vote only matters in the event the voter is pivotal, conditioning on this event generally reveals information about the state through the strategies of other voters \citep{Au1990}. This effect is captured by $\mathbb P_i(q-1\mid\omega)$ in \eqref{difference1} and also depends on the precision of private signals. However, it is not apparent that such dependence disappears when the signals become perfectly precise.

In Lemma \ref{benchmark:between}, I show that the voters with $q$ highest ideal policies vote to accept any policy between $y^\ell_q$ and $y^h_q$ after receiving high signals when signals are sufficiently precise, implying that such policies are certainly accepted in state $h$. Further, I show that the voters with $(N-q+1)$ lowest ideal policies vote to reject any policy between $y^\ell_q$ and $y^h_q$ after receiving low signals with probability approaching one as signals become perfectly precise, implying that such policies are certainly rejected in state $\ell$. Monotonicity of voting strategies is used to show that each voter is either never pivotal or there is a strict probability of being pivotal in state $h$. Moreover, the properties of Poisson binomial random variables allow us to place a bound on the limit ratio of pivotal probabilities along the sequences of strategy profiles.\footnote{Poisson random variables and their properties are discussed in Appendix \ref{appendix-prelim}.}

Given the voting behavior described above, the agenda-setter's choice essentially boils down to policies $y^\ell_q$ and $y^h_q$.\footnote{More precisely, the choice is between policy $y^\ell_q$ and a policy that maximizes the agenda-setter's payoff in state $h$. Such policy necessarily converges to $y^h_q$.} With probability approaching one: policy $y^\ell_q$ is accepted in both states, and policy $y^h_q$ is accepted in state $h$ and rejected in state $\ell$. The agenda-setter's choice depends on the prior belief that the state is $h$ and, in particular, whether this prior probability is high enough to justify the risk of implementing a status-quo policy. The following benchmark formalizes this result.

\begin{bench}[Take-it-or-leave-it offer]\label{bench:take}
  Fix the required quota $q$ and a sequence of signal precisions $\{\tau^n\}$ converging to 1. The agenda-setter's expected payoff converges to $V^T_A(\mu^0)=\max\{y^\ell_q,\mu^0(h)y^h_q\}$.
\end{bench}

Notably, Lemmas \ref{benchmark:belowabove} and \ref{benchmark:between} claim that a fixed committee with a $q$-majority voting rule fully aggregates the private information of voters. For a given proposal $p$, the probability that the voting outcome is the same as when the private signals are publicly revealed converges to one as the signals become perfectly precise. This statement holds for all proposals except $y^h_q$, in which case there exists a sequence of equilibria in which information is not fully aggregated, although there also exists one in which it is. Such interpretation complements the results on information aggregation by large committees with private signals of a fixed precision, as discussed in the introduction.

\section{Main results}\label{sec-main}

Finding the agenda-setter's highest expected payoff requires two steps. First, an upper bound is placed on the agenda-setter's expected payoff, and then it is shown that this upper bound is tight, meaning that it can be achieved in equilibrium. The second step is constructive and follows the ideas presented in Section \ref{sec-ideas}.\footnote{Additional details are provided in the proofs of Theorems \ref{theorem-equal} and \ref{theorem-unequal}.} The first step is more complicated and is the focus of the discussion in this section. To show that the agenda-setter's expected payoff cannot exceed some level, it is sufficient to show that specific policy sequences cannot be implemented with a positive probability in any equilibrium. In turn, this amounts to showing that in any putative equilibrium that induces such policy sequences, there must exist some player with a profitable deviation. For our purposes, it is not necessary to consider the proposal strategy and sufficient to find a voter with a profitable deviation.

I begin with the case when voters observe private signals with the same precision and then consider the case when one voter has an asymptotically more precise signal than other voters. Together, these cases demonstrate the range of possible outcomes and various considerations that might arise in the general case.

\begin{theo}\label{theorem-equal}
  Let $\{\tau_i^n\}_{n=1}^\infty=\{\tau^n\}_{n=1}^\infty\to1$ for all $i\in\mathcal N$ and $q\geq2$. The limit of the agenda-setter's highest expected payoff as private signals become perfectly precise and the players become perfectly patient equals $V_A(\mu^0)=\mu^0(h)y^h_q+\mu^0(\ell)y^\ell_q$, i.e., the agenda-setter achieves the complete information benchmark.
\end{theo}

The fact that $\mu^0(h)y^h_q+\mu^0(\ell)y^\ell_q$ is an upper bound follows from Lemmas \ref{lem:high} and \ref{lem:high2}, which state that the probability that any policy greater than $y^\omega_q$ is accepted in state $\omega$ must converge to zero. This result is similar to Lemmas \ref{benchmark:belowabove} and \ref{benchmark:between} used in the proof of Benchmark \ref{bench:take}, but requires accounting for the signaling incentives of voters and the continuation payoffs in case the proposal is rejected. The difference in the expected payoff of voter $i$ with signal $s_i$ and belief $\mu$ from voting to accept and reject policy $p$ can be written as:
\begin{align}
  \mathbb EU_i(p;s_i,\mu,\alpha_{-i}) & \propto \sum_{\omega\in\{\ell,h\}}\mu(\omega)\tau_i^\omega(s_i)\Big\{ \mathbb P_i(q-1\mid\omega) [u_i^\omega(p)-u_i^\omega(0)] \nonumber\\ 
  & +\sum_{a_{-i}\in A^{q-1}}\delta\mathbb P_i(a_{-i}\mid\omega) \left[u_i^\omega(0)-V_i(s_i,\tilde\mu(\mu;p,(a_{-i},0)))\right] \nonumber\\
  & +\sum_{\substack{a_{-i}\in A^r \\ r\leq q-2}}\delta\mathbb P_i(a_{-i}\mid\omega) \big[V_i(s_i,\tilde\mu(\mu;p,(a_{-i},1))) - V_i(s_i,\tilde\mu(\mu;p,(a_{-i},0)))\big] \Big\}, \label{difference3}
\end{align}
where for each $r\in\mathcal N$, $A^r$ is the set of voting profiles $a_{-i}$ of voters other than $i$ such that the number of voters who vote to accept the proposal equals $r$. The expression \eqref{difference1} used in the proof of Benchmark \ref{bench:take} only contains the terms multiplied by the probability $\mathbb P_i(q-1\mid \omega)$ of being pivotal and replaces the continuation payoff $V_i(s_i,\tilde\mu(\mu;p,a))$ with the status-quo payoff $u_i^\omega(0)$. In contrast, \eqref{difference3} accounts for signaling incentives and the fact that pivotal incentives depend on continuation payoff rather than the status-quo payoff. 

The proofs of Lemmas \ref{lem:high} and \ref{lem:high2} use the properties of Poisson binomial random variables to place a bound on the limit ratios of pivotal and signaling probabilities along the sequences of strategy profiles. In fact, focusing on skimming equilibria allows us to show that for every voter in $\{q,\dots,N\}$, the pivotal and signaling incentives are aligned. For state $\ell$, we only need to notice that the support of implemented policies is bounded below by $y^\ell_q$. The argument for state $h$ is more involved because the pivotal and signaling incentives of these voters may not be aligned. In particular, the pivotal incentive of voter $i\in\{q,\dots,N\}$ is to reject $p>y^h_q$ in favor of smaller implemented policies. However, the signaling incentive depends on whether voter $i$ prefers a first-order stochastically dominated distribution of implemented policies, which is not necessarily the case and depends on the ideal policy of voter $i$ in state $h$ and the induced distributions of policies. Nonetheless, I show that the support of implemented policies in state $h$ is bounded below by $\max\{y^h_q-y^\ell_q,y^\ell_q\}$, which implies that the pivotal and signaling incentives of voters in $\{q,\dots,N\}$ are aligned in that state as well.

As mentioned in Section \ref{sec-ideas}, an asymptotically more precise signal dominates other signals in determining posterior belief when revealed by voting strategies. On the one hand, this means that voters ignore their private signals when conditioning on the event that reveals the signal of the more informed voter. On the other hand, this means that the more informed voter ignores the information revealed by other players and solely influences the posterior belief about the state. The following result shows that such a voter disproportionately influences the agenda-setter's ability to extract surplus from the decisive voter.

\begin{theo}\label{theorem-unequal}
  Let $\{\tau_j^n\}_{n=1}^\infty\to1$ for all $j\in\mathcal N$ and suppose there is $i\leq q$ such that $\lim_{n\to\infty}\frac{1-\tau_i^n}{\prod_{j\neq i}(1-\tau_j^n)}=0$. The limit of the agenda-setter's highest expected payoff as private signals become perfectly precise and the players become perfectly patient equals $V_A(\mu^0)=\mu^0(h)\max\{y^\ell_q,\min\{y^h_q,y^h_i-y^\ell_q\}\}+\mu^0(\ell)y^\ell_q$.
\end{theo}

Intuitively, it is easy to see why the agenda-setter's expected payoff in state $h$ can be at most $(y^h_i-y^\ell_q)$ when this policy is greater than $y^\ell_q$. Suppose $p>y^h_i-y^\ell_q>y^\ell_q$ is accepted in state $h$ with probability greater than some positive threshold $\varepsilon$ for very precise signals. If the putative equilibrium has voter $i$ use an uninformative voting strategy, she will vote as if being pivotal while ignoring any possible information revealed by other voters. Given the assumed ranking of ideal policies, voter $i$ will vote to accept $p$ after receiving a high signal and reject $p$ after receiving a low signal. Therefore, she must use an informative voting strategy, implying that the revised policy after voter $i$'s rejection is $y^\ell_q$. But then she will vote to reject $p$ today after receiving a high signal to implement policy $y^\ell_q$ tomorrow, which is a contradiction.

What if a voter $i>q$ has an asymptotically more precise signal than other voters? It turns out that the signal precision of voters with $N-q$ lowest ideal policies has no effect on the agenda-setter's highest expected payoff. In particular, an upper bound in Lemma \ref{lem:bounds} is derived for any signal structure, and the proof of Theorem \ref{theorem-equal} continues to hold when only voters with $q$ highest ideal policies receive equally precise signals.

The following corollary summarizes the necessary and sufficient conditions for the agenda-setter to achieve the complete information benchmark and for the Coase conjecture to hold. Overall, the conflict in preferences between voter $i$ in state $h$ and voter $q$ in state $\ell$, captured by the difference $(y^h_i-y^\ell_q)$, plays a crucial role. The Coase conjecture holds when this conflict is sufficiently small, and the complete information benchmark can be achieved when this conflict is sufficiently large. In the intermediate case, the Coase conjecture is violated, but the complete information benchmark cannot be achieved, meaning that the agenda-setter can extract some surplus from voter $q$ in state $h$ but not all of it.

\begin{cor}\label{corollary}
  Suppose voter $i\leq q$ has an asymptotically more precise signal than other voters.
  \begin{enumerate}[(a)]
    \item The agenda-setter can achieve the complete information benchmark if and only if $(y^h_i-y^\ell_q)\geq y^h_q$.
    \item The Coase conjecture holds if and only if $(y^h_i-y^\ell_q)\leq y^\ell_q$.
  \end{enumerate}
\end{cor}

\subsection{Change in the required quota}\label{sec-quota}

An increase in the required quota implies that the ideal policies of the target voter decrease and the agenda-setter must craft proposals acceptable to voters with smaller ideal policies, harming the agenda-setter in a static context (see Benchmarks \ref{bench:full} and \ref{bench:take}). Surprisingly, when the information about the state is dispersed among voters and the agenda-setter can make revised proposals, she may benefit from an increased threshold. Perhaps the most interesting case when this happens is when voter $i\leq q$ has a more informative signal than other voters.\footnote{There are other cases as well. First, it is easy to derive sufficient conditions for a case with an increase in threshold from the initial level $q=1$ under the conditions of Theorem \ref{theorem-equal}, namely, equal precision of signals. Second, one can consider a case when voter $q$ has an asymptotically more precise signal than voters in $\{1,\dots,q-1\}$ and at the same time, voter $\tilde q>q$ has an asymptotically more precise signal than voters in $\{1,\dots,\tilde q-1\}$. The sufficient conditions for an increase in threshold from $q$ to $\tilde q$ can be derived using Theorem \ref{theorem-unequal}.} Suppose that $y^h_q>y^h_i-y^\ell_q>y^\ell_q$, so that the agenda-setter can extract some surplus from voter $q$ in state $h$, but not all of it (see Figure \ref{util}). Then, the agenda-setter's highest expected payoff increases in response to an increase in the voting threshold from $q$ to $\tilde q$ when the prior belief $\mu(h)$ is greater than $\frac12$ when $(y^h_i-y^\ell_{\tilde q})\leq y^h_{\tilde q}$ or greater than $\frac{y^\ell_q-y^\ell_{\tilde q}}{2y^\ell_q+y^h_{\tilde q}-y^h_i-y^\ell_{\tilde q}}$ when $(y^h_i-y^\ell_{\tilde q})>y^h_{\tilde q}$.

This example highlights the role of the voting threshold in providing incentives for voters to accept policies in screening equilibria. As shown in the proof of Theorem \ref{theorem-unequal}, the agenda-setter makes finitely many screening proposals that converge to policy $p=(y^h_i-y^\ell_q)$ that makes voter $i$ in state $h$ indifferent between $p$ and $y^\ell_q$. If all of these proposals are rejected, the agenda-setter proposes $y^\ell_q$. An increase in the voting threshold from $q$ to $\tilde q$ decreases the anticipated revised proposal from $y^\ell_q$ to $y^\ell_{\tilde q}$. But since voter $i$ has single-peaked preferences and $\frac 12y^h_i>y^\ell_q$, she prefers $y^\ell_q$ to $y^\ell_{\tilde q}$ in state $h$. Therefore, the screening policy $\tilde p$ that makes her indifferent between $\tilde p$ and $y^\ell_{\tilde q}$ is, in fact, greater than $p$, implying that the agenda-setter can extract more surplus from voter $q$ in state $h$. The argument is completed by noticing that the prior belief that the state is $h$ needs to be sufficiently strong so that an increase in implemented policy in state $h$ is not offset in the calculation of expected payoff by a decrease in implemented policy in state $\ell$.

\subsection{Ability to make revised proposals}

Our main results allow us to describe the circumstances when the ability to make revised proposals is valuable (or harmful) to the agenda-setter. Recall from Section \ref{bench-take} that when the agenda-setter makes a single proposal, she can extract all surplus from voter $q$ in state $h$ at the cost of implementing the status quo policy 0 in state $\ell$. In contrast, when the agenda-setter can make revised proposals, she can extract some surplus from voter $q$ in state $h$ while implementing a policy greater than the status quo in state $\ell$, namely, a policy close to $y^\ell_q$. To summarize the conditions under which the ability to make revised proposals is valuable or harmful to the agenda-setter, write the agenda-setter's highest expected payoff as $V_A(\mu^0)=y^\ell_q+\mu^0(h)(p^{**}-y^\ell_q)$, where $p^{**}$ takes a value in $\{y^\ell_q,y^h_q,y^h_i-y^\ell_q\}$ depending on the relative precision of private information and the conflict in preferences among voters and between states.\footnote{Notice that we restrict our attention to cases characterized in Theorems \ref{theorem-equal} and \ref{theorem-unequal}.} In turn, the agenda-setter's highest expected payoff without the ability to make revised proposals can be written as $V_A^T(\mu^0)=y^\ell_q+\max\{0,\mu^0(h)y^h_q-y^\ell_q\}$ (see Benchmark \ref{bench:take}). Some examples are provided in Figure \ref{fig-coasian}. Notice that $V_A(\mu^0)$ is a linear function of the prior belief $\mu^0(h)$ and its non-negative slope depends on the agenda-setter's ability to extract surplus from voter $q$ in state $h$. However, the slope can take any value in $[0,y^h_q]$, depending on the environment, and equals 0 when the Coase conjecture holds. At the same time, $V_A^T(\mu^0)$ is piecewise linear in $\mu^0(h)$ and has a kink at $\mu^0(h)=y^\ell_q/y^h_q$. However, its slope is strictly positive and equals $y^h_q$ for sufficiently high prior belief that the state is $h$.

The precise conditions for ranking $V_A(\mu^0)$ and $V^T_A(\mu^0)$ are given in Appendix \ref{appendix-ability}. Overall, the ability to make revised proposals is valuable to the agenda-setter when she can extract all surplus from voter $q$ in state $h$, or when she can extract some surplus and the prior belief that the state is $h$ is sufficiently weak. In contrast, the ability to make revised proposals is harmful to the agenda-setter when she cannot extract all surplus and the prior belief that the state is $h$ is sufficiently strong.

\section{Related literature and directions for future work}\label{sec-lit}

Besides the related work mentioned in the introduction, this paper contributes to the literature on dynamic decision-making through voting; in particular, the literature on repeated referenda originated from \citet*{RoRo79}.\footnote{\citet*{RoRo79} builds on a single-period model of \citet*{RoRo78}. \cite{DeMa83} is another influential paper with a uniformed agenda-setter making a single proposal to privately informed voters.} The first paper to introduce strategic voting in the model of repeated referenda was \citet*{Mo88}.\footnote{Even though \cite{Mo88} considers strategic voters and discusses the potential conflict between the signaling and pivotal incentives, that paper eventually assumes that the pivotal incentives dominate.} More recent work on repeated referenda includes \cite*{cameron2000veto}, \cite*{primo2002rethinking}, and \cite{barseghyan2014bureaucrats}. Most closely related papers are \citet*{rosenthal2022sequential} and \citet*{chen2023sequential}, which consider two-period models without time discounting. \citet*{rosenthal2022sequential} consider a model with a single voter and focus on the effect of voter sophistication on the agenda-setter's ability to benefit from the option to revise the initial proposal. \citet*{chen2023sequential} allows for multiple voters, introducing the possibility of signaling when voting on the initial proposal. In contrast to these papers, my model allows for infinite periods and time discounting, which creates intertemporal incentives and allows us to study the ability of the agenda-setter to screen private information. I also assume that voters have correlated preferences and learn about the common state from each other's voting behavior and, more subtly, that players use Markov strategies, making the results more robust to ad hoc specification of behavior off the equilibrium path.

This paper also contributes to the literature studying the role of voting schemes in the presence of asymmetric information and agenda control (see \cite*{AustinSmith} for an early example). \cite{BoEr2010} study the effect of the voting rule on the efficiency of an adopted policy when the agenda-setter proposes the policy that is voted on. \citet*{Bo2021} compare the efficiency of ``voting mechanisms'' when the agenda-setter serves as a gatekeeper and decides whether the vote takes place. Both papers focus on the aggregation of information dispersed among voters and allow the agenda-setter to be privately informed but assume that no revisions occur when a policy does not gather enough support. In contrast, the central features of our analysis are the agenda-setter's ability to revise rejected proposals and the associated changes in players' incentives. Moreover, both papers consider common value environments,\footnote{More precisely, \cite{BoEr2010} assume that the preferences of voters are almost perfectly aligned and \citet*{Bo2021} allow partisan voters whose preferences do not depend on the underlying state of the world.} while I emphasize the role of heterogeneity in preferences among voters. \cite{He08} studies bargaining over a distributive policy and shows that the agenda-setter may offer positive transfers to more voters than the required quota. I focus on a one-dimensional policy space that does not allow the agenda-setter to make targeted transfers to voters.

This paper is also related to the literature on the role of voting rule in collective search \citep*{Albrecht,CompteJehiel,MoldovanuShi}, collective experimentation \citep*{Strulovici}, and sequential voting with private information \citep*{Ordeshook,KleinerMoldovanu}. Policy proposals arrive exogenously in these papers until the committee collectively decides to stop the search and accept the current proposal. In contrast, I assume that the agenda-setter endogenously selects the policy proposals.

Multiple avenues exist for future work on multilateral bargaining with private information or single-peaked preferences. The techniques in this paper can be used to analyze a model in which the agenda-setter's utility is state-dependent, similar to \cite*{DL}. A vital but more challenging extension is to allow players with private information to make proposals.\footnote{\citet*{Ba90,Ba93} and \citet*{Lupia} study single-period models of monopoly agenda control with a privately informed agenda-setter and focus on the signaling role of proposals.}

\bibliographystyle{ecta}
\bibliography{cites-finite}

\begin{appendix}

\section{Preliminaries}\label{appendix-prelim}

A discrete random variable $Y$ follows a Poisson binomial distribution if $Y=\sum_{i=1}^NX_i$ where $\{X_i\}_{i=1}^N$ are independent Bernoulli random variables. The distribution of $Y$ is completely characterized by the vector of success probabilities $z=(z_1,\dots,z_N)$. The Binomial distribution is a special case when $z_i=z_j$ for all $i,j=1,\dots,N$. A Poisson binomial random variable $Y$ is supported on $\{0,\dots,N\}$ and has a probability mass function
\begin{equation}\label{poisson-pmf}
  f(r\mid z)=\sum_{C\in\mathcal N^r}\prod_{i\in C}z_i\prod_{j\not\in C}(1-z_j),  
\end{equation}
where $\mathcal N^r$ is the collection of all subsets of $\mathcal N=\{1,\dots,N\}$ of size $r\in\{0,\dots,N\}$. This definition immediately implies that we can write for each $i\in\mathcal N$ and $r\in\{1,\dots,N\}$:
\begin{equation}\label{poisson-decomp}
f(r\mid z)=z_i f(r-1\mid z_{-i})+(1-z_i)f(r\mid z_{-i}).
\end{equation}
For this section only, let $l(f(\cdot\mid z))$ and $m(f(\cdot\mid z))$ be the lowest and the highest mode of the distribution $f(\cdot\mid z)$. Of course, it is possible that $l(f(\cdot\mid z))=m(f(\cdot\mid z))$. In addition, define $\mu=\sum_{i\in\mathcal N}z_i$ and let $\delta(z)=\mu-[\mu]$ to be the fractional part of $\mu$.\footnote{The notation in this section is kept closely in line with the notation in \cite{darroch1964distribution} and \cite{samuels1965number}. For instance, $[\cdot]$ is defined as $[x]=\max\{y\in\mathbb Z\mid y\leq x\}$ for all $x\in\mathbb R$.} We have the following characterization of the distribution of $Y$.
\begin{lem}[\cite{darroch1964distribution,samuels1965number}]\label{ref:poisson} 
  Let $Y$ be a Poisson binomial random variable with success probabilities $z$ and pmf $f(\cdot\mid z)$. Then:
  \begin{enumerate}[(a)]
    \item The pmf $f(\cdot\mid z)$ has at most two modes. Moreover, letting $\delta^-=1/(N+1)$ and $\delta^+=N/(N+1)$, we have:
    \[
      \mathrm{Mode}(f(\cdot\mid z)) = \begin{cases}
        \mu & \text{ if } \delta(z) = 0, \\
        [\mu] & \text{ if } \delta(z)\in(0,\delta^-), \\
        \text{$[\mu]$ or $[\mu]+1$ or both} & \text{ if } \delta(z)\in(\delta^-,\delta^+), \\
        [\mu]+1 & \text{ if } \delta(z)\in(0,\delta^+).
      \end{cases}
    \]
    \item The pmf $f(\cdot\mid z)$ strictly increases on $\{0,\dots,l(f(\cdot\mid z))\}$ and strictly decreases on $\{m(f(\cdot\mid z)),\dots,N\}$.
  \end{enumerate}
\end{lem}
In particular, Lemma \ref{ref:poisson} implies that the distribution of $Y$ is bell-shaped and has at most two modes that are consecutive integers. Moreover, the mode of $Y$ is uniquely pinned down when $\mu=\sum_{i\in\mathcal N}z_i$ is close to an integer.

The following result compares the pmf of two Poisson binomial random variables supported on $\{1,\dots,N-1\}$ with success probabilities $z$ and $z'$ such as $z\leq z'$.

\begin{lem}\label{ranking:key}
  Fix $q\in\mathcal N$, $\varepsilon\in(0,1/(N+1))$, and assume $z_i\leq z'_i$ for all $i\in\mathcal N$.
  \begin{enumerate}[(a)]
    \item If $\sum_{i=1}^N z_i\geq q-\varepsilon$, then $f(q-1\mid z)\geq f(q-1\mid z')$.
    \item If $\sum_{i=1}^N z'_i\leq q-1+\varepsilon$, then $f(q\mid z)\leq f(q\mid z')$.
  \end{enumerate}
\end{lem}
\begin{proof}[Proof of Lemma \ref{ranking:key}]
  Since $z\leq z'$ point-wise, we can obtain $z'$ from $z$ by a series of $N$ steps, weakly increasing $z_k$ by $\Delta_k=(z'_k-z_k)$ at $k$-th step. Let $z^k$ be the resulting vector of success probabilities after $k$-th step so that $z^0=z$ and $z^{N}=z'$. For $r\in\mathcal N$, after $k$-th step we have:
  \begin{align}\label{diff:poisson}
  f(r\mid z^k)-f(r\mid z^{k-1})&=z^k_k f(r-1\mid z^{k-1}_{-k})+(1-z^k_k)f(r\mid z^{k-1}_{-k}) \nonumber\\
  &-z^{k-1}_k f(r-1\mid z^{k-1}_{-k})-(1-z^{k-1}_k)f(r\mid z^{k-1}_{-k}) \nonumber\\
  &=\Delta_k f(r-1\mid z^{k-1}_{-k})-\Delta_k f(r\mid z^{k-1}_{-k}) \nonumber\\
  &=\Delta_k \big[f(r-1\mid z^{k-1}_{-k})-f(r\mid z^{k-1}_{-k})\big],
  \end{align}
  where $\Delta_k\geq 0$ and the first equality uses \eqref{poisson-decomp} and the fact the $z^k$ and $z^{k-1}$ differ only in $k$-th component so $z^k_{-k}=z^{k-1}_{-k}$.
  
  \textit{(a)} Notice that $\sum_{i\neq k} z^{k-1}_i\geq \sum_{i\neq k} z_i$ by construction; moreover, $\sum_{i\neq k} z_i\geq q-1-\varepsilon$ since $\sum_{i=1}^N z_i\geq q-\varepsilon$ by assumption. Therefore, $\sum_{i\neq k} z^{k-1}_i\geq q-1-\varepsilon$. By the choice of $\varepsilon$, the smallest mode of $f(\cdot\mid z^{k-1}_{-k})$ is at least $(q-1)$ by part \textit{(a)} of Lemma \ref{ref:poisson} and therefore $f(q-2\mid z^{k-1}_{-k})<f(q-1\mid z^{k-1}_{-k})$ by part \textit{(b)} of Lemma \ref{ref:poisson}. Letting $r=q-1$ in \eqref{diff:poisson}, if follows that $f(q-1\mid z^k)-f(q-1\mid z^{k-1})\leq 0$ for all $k=1,\dots,N$ and therefore $f(q-1\mid z')\leq f(q-1\mid z)$, which is the desired result.

  \textit{(b)} Similar to part \textit{(a)}, we can see that $\sum_{i\neq k} z^{k-1}_i\leq \sum_{i\neq k} z'_i\leq q-2+\varepsilon$ and the largest mode of $f(\cdot\mid z^{k-1}_{-k})$ is at most $(q-1)$, which implies $f(q-1\mid z^{k-1}_{-k})>f(q\mid z^{k-1}_{-k})$. Letting $r=q$ in \eqref{diff:poisson}, if follows that $f(q\mid z^k)-f(q\mid z^{k-1})\geq 0$ for all $k=1,\dots,N$ and therefore $f(q\mid z')\geq f(q\mid z)$, which is the desired result.
\end{proof}

We will use Lemma \ref{ranking:key} in a slightly different form. Fixing a vector of voting strategies $\alpha$, notice th t for any voter $i\in\mathcal N$, the number of votes by other voters for a given proposal $p$ (given belief $\mu$) is a Poisson binomial random variable. For instance, conditional on all other voters receiving the same signal $s\in\{\rl,\rh\}$, the vector of success probabilities is $\alpha_{-i}(p;s,\mu)$. And since the voting strategies are monotone in signal, we have $\alpha_{-i}(p;\rl,\mu)\leq \alpha_{-i}(p;\rh,\mu)$. Now, suppose that there is a sequence of voting strategy profiles $\{\alpha^n\}_{n=1}^\infty$ such that $\sum_{j\neq i}\alpha_{j}^{n}(p;\rl,\mu)\geq q$. Then, for a sufficiently high $n$ we have $\sum_{j\neq i}\alpha_{j}^{n}(p;\rl,\mu)\geq q-1/N$. Now, we can apply Lemma \ref{ranking:key} to $z=\alpha^{n}_{-i}(p;\rl,\mu)$ and $z'=\alpha^{n}_{-i}(p;\rh,\mu)$.

\begin{lem}\label{ranking:greater}
  Let $p\in[0,M]$ and $\{\sigma^n\}_{n=1}^\infty$ be a sequence of strategy profiles with monotone voting strategies and suppose $\lim_{n\to\infty}\sum_{j\neq i}\alpha^{n}_j(p;\rl,\mu)\geq q$. Then, $\mathbb P_i^n(q-1\mid\ell)\geq\mathbb P_i^n(q-1\mid h)$ for all $i\in\mathcal N$ and $n$ sufficiently large.
\end{lem}
In turn, if there is a sequence of voting strategy profiles $\{\alpha^n\}_{n=1}^\infty$ such that $\sum_{j\neq i}\alpha_{j}^{n}(p;\rh,\mu)\leq q-2+\varepsilon$, then Lemma \ref{ranking:key} implies the next result. Together, Lemmas \ref{ranking:greater} and \ref{ranking:lower} will help us establish the asymptotic bounds on the signaling incentives of voters.
\begin{lem}\label{ranking:lower}
  Let $p\in[0,M]$ and $\{\sigma^n\}_{n=1}^\infty$ be a sequence of strategy profiles with monotone voting strategies and suppose $\lim_{n\to\infty}\sum_{j\neq i}\alpha^{n}_j(p;\rh,\mu)\leq q-2$. Then, $\mathbb P_i^n(q-1\mid\ell)\leq\mathbb P_i^n(q-1\mid h)$ for all $i\in\mathcal N$ and $n$ sufficiently large.
\end{lem}

The following Lemmas allow us to establish a connection between the probability of acceptance and individual voting strategies for precise signals. We only prove Lemma \ref{bound:high} since the proof of Lemma \ref{bound:low} is analogous.

\begin{lem}\label{bound:high}
  Take a sequence $\{(\tau^n_i)_{i\in\mathcal N}\}_{n=1}^\infty$ such that $\tau^n_i$ converges to 1 for each $i\in\mathcal N$ and let $\{\sigma^n\}_{n=1}^\infty$ be a sequence of strategy profiles. Fix $p\in[0,M]$ and suppose that $\{\sum_{r\geq q}\mathbb P^n(r\mid h)\}_{n=1}^\infty$ does not converge to 1. Then, there are at least $(n-q+1)$ players $i\in\mathcal N$ such that $\{\alpha_i^{n}(p;\rh,\mu)\}_{n=1}^\infty$ does not converge to 1.
\end{lem}
\begin{proof}[Proof of Lemma \ref{bound:high}]
  By assumption, there exists $\varepsilon>0$ such that for each $m$ there is $n\geq m$ for which $\sum_{r\geq q}\mathbb P^n(r\mid h)\leq 1-\varepsilon$. We want to show that there is $C\subseteq\mathcal N$ such that $\lvert C\rvert\geq n-q+1$ and for each $i\in C$ there is $\varepsilon_i>0$ with the property that for each $m$ there exists $n\geq m$ for which $\alpha_i^{n}(p;\rh,\mu)\leq 1-\varepsilon_i$.

  The proof is by contradiction. Denote $C^*=\{i\in\mathcal N\mid \lim_{n\to\infty}\alpha_i^{n}(p;\rh,\mu)=1 \}$ and suppose $\lvert C^*\rvert\geq q$. To establish a contradiction, it is sufficient to show that $\{\sum_{r<q}\mathbb P^n(r\mid h)\}_{n=1}^\infty$ converges to 0, because then $\{\sum_{r\geq q}\mathbb P^n(r\mid h)\}_{n=1}^\infty$ converges to 1. Rewrite $ \sum_{r<q}\mathbb P^n(r\mid h)$ as $\sum_{r<q}\mathbb P^n(r\mid h)=\sum_{r<q}f(r\mid (\tau^n\alpha_i^{n}(p;\rh,\mu)+(1-\tau^n)\alpha_i^{n}(p;\rl,\mu))_{i\in \mathcal N})$, where $f(r\mid p)$ is the Poisson binomial pmf defined in \eqref{poisson-pmf}. Now, notice that for each $C\in\mathcal N^r=\{C\subset\mathcal N\mid \lvert C\rvert=r\}$ for $r<q$, there exists $i\in C^*$ such that $i\not\in C$. But this immediately implies that $\{f(r\mid (\tau_i^n\alpha_i^{n}(p;\rh,\mu)+(1-\tau_i^n)\alpha_i^{n}(p;\rl,\mu))_{i\in\mathcal N})\}_{n=1}^\infty$ converges to 0. Since this conclusion holds for each $r<q$, we obtain that $\{\sum_{r<q}\mathbb P^n(r\mid h)\}_{n=1}^\infty$ converges to 0, which completes the proof.
\end{proof}

\begin{lem}\label{bound:low}
  Take a sequence $\{(\tau^n_i)_i\}_{n=1}^\infty$ such that $\tau^n_i$ converges to 1 for each $i\in\mathcal N$ and let $\{\sigma^n\}_{n=1}^\infty$ be a sequence of strategy profiles. Fix $p\in[0,M]$ and suppose that $\{\sum_{r\geq q}\mathbb P^n(r\mid \ell)\}_{n=1}^\infty$ does not converge to 0. Then, there are at least $q$ players $i\in\mathcal N$ such that $\{\alpha_i^{n}(p;\rl,\mu)\}_{n=1}^\infty$ does not converge to 0.
\end{lem}

\section{Benchmark 2: Take-it-or-leave-it offer}\label{appendix-bench2}

Consider the difference in the expected payoffs of voter $i$ from voting to accept and voting to reject proposal $p$ given signal $s_i$. If the probability that $i$ is pivotal is positive, $\mathbb P_i(q-1)>0$ (where the dependence on the proposal and voting strategies is implicit), then the difference in the expected payoffs, denoted by $\mathbb EU_i(p;s_i,\mu,\alpha_{-i})$, is given by \eqref{difference1}, which uses the conditional independence of signals. In \eqref{difference1}, the left-hand side equals the right-hand side multiplied by the unconditional probability that voter $i$ observes signal $s_i$. Since this probability is strictly positive and does not depend on other players' strategies, it can be ignored.

If the probability that $i$ is pivotal equals 0, then the difference in the expected payoffs is also 0, and voter $i$ votes sincerely; that is, voter $i$ votes based on the difference in the expected payoffs written as follows:
\begin{equation}\label{difference2}
  \mathbb EU_i(p;s_i,\mu) \propto \sum_{\omega\in\{\ell,h\}}\mu(\omega)\tau_i^\omega(s_i)(u_i^\omega(p)-u_i^\omega(0)).
\end{equation}

Consider a sequence $\{(\tau_i^n)_i\}$ such that $\tau_i^n$ converges to 1 for each $i\in\mathcal N$ and let $\{\sigma^n\}$ be corresponding sequence of equilibrium strategy profiles. For each $n\geq1$, each player $i\in\mathcal N$, and each signal $s_i\in S=\{\rl,\rh\}$, we have:
\begin{align}
  \alpha_i(p;s_i,\mu) & =
  \begin{cases}
    1 & \text{ if } \mathbb EU_i^n(p;s_i,\mu,\alpha^n_{-i})>0 \\
    0 & \text{ if } \mathbb EU_i^n(p;s_i,\mu,\alpha^n_{-i})<0 
  \end{cases} & \text{ when } \mathbb P_i(q-1)>0, \text{ and} \label{cond:alpha1} \\
  \alpha_i(p;s_i,\mu) & =
  \begin{cases}
    1 & \text{ if } \mathbb EU_i^n(p;s_i,\mu)>0 \\
    0 & \text{ if } \mathbb EU_i^n(p;s_i,\mu)<0 
  \end{cases} & \text{ when } \mathbb P_i(q-1)=0. \label{cond:alpha2}
\end{align}

These equilibrium restrictions on the voting strategy of voter $i$ immediately imply that proposals below $y^\ell_q$ are accepted with certainty and proposals above $y^h_q$ are rejected with certainty, as stated in the following lemma.

\begin{lem}\label{benchmark:belowabove}
  Fix any sequence $\{\tau^n\}_{n=1}^\infty$ and an associated sequence of strategy profiles $\{\sigma^n\}$. Then, for each state $\omega\in\{\ell,h\}$:
  \begin{enumerate}[(a)]
    \item If $p<y^\ell_q$, then $\{\sum_{r\geq q}\mathbb P^n(r\mid \omega)\}_{n=1}^\infty$ is constant and equals 1.
    \item If $p>y^h_q$, then $\{\sum_{r\geq q}\mathbb P^n(r\mid \omega)\}_{n=1}^\infty$ is constant and equals 0.
  \end{enumerate}
\end{lem}
\begin{proof}[Proof of Lemma \ref{benchmark:belowabove}]
  Consider voter $i\in\{1,\dots,q\}$. Since $p<y^\ell_q$, \eqref{cond:alpha1} and \eqref{cond:alpha2} imply that $\alpha_i^{n}(p;\rl,\mu)=\alpha_i^{n}(p;\rh,\mu)=1$ for each $n\geq1$. And since at least $q$ voters accept $p$ with certainty, the probability that $p$ gets at least $q$ votes equals 1. Similarly, \eqref{cond:alpha1} and \eqref{cond:alpha2} imply that $\alpha_i^{n}(p;\rl,\mu)=\alpha_i^{n}(p;\rh,\mu)=0$ for each $i\in\{q,\dots,N\}$ and each $n\geq1$ when $p>y^h_q$. Since at least $(N-q+1)$ voters reject $p$ with certainty, the probability that $p$ gets at least $q$ votes equals 0.
\end{proof}

Establishing facts about the probability of acceptance of proposal $p\in(y^\ell_q,y^h_q)$ is more challenging. The first issue is that even when strategic voters end up voting sincerely and informatively, collective mistakes are possible due to the limited precision of signals. This problem is alleviated by considering a sequence of equilibria as the precision of signals converges to one. The second issue is that when the probability of being pivotal is positive, voters receive equilibrium information that may override their private signals and lead to uninformative voting by at least some voters.

Despite these issues, the following result shows that the probability that policy $p$ between $y^\ell_q$ and $y^h_q$ is accepted converges to 1 when the state is $h$ and 0 when the state is $\ell$.
\begin{lem}\label{benchmark:between}
  Take a sequence $\{\tau_n\}$ converging to 1 and let $p\in(y^\ell_q,y^h_q)$. Then:
  \begin{enumerate}[(a)]
    \item $\{\sum_{r\geq q}\mathbb P^n(r\mid h)\}_{n=1}^\infty$ converges to 1.
    \item $\{\sum_{r\geq q}\mathbb P^n(r\mid \ell)\}_{n=1}^\infty$ converges to 0.
  \end{enumerate} 
\end{lem}
\begin{proof}[Proof of Lemma \ref{benchmark:between}]\hfill

  \textit{(a)} Suppose $\{\sum_{r\geq q}\mathbb P^n(r\mid \ell)\}_{n=1}^\infty$ does not converge to 0, so there exists $\varepsilon>0$ such that for each $m$ there is $n\geq m$ for with $\sum_{r\geq q}\mathbb P^n(r\mid \ell)>\varepsilon$. By Lemma \ref{bound:low}, there are at least $q$ players $i$ for which there exists $\varepsilon_i>0$ such that for each $m$ there is $n\geq$ with the property that $\alpha_i^{n}(p;\rl,\mu)>\varepsilon_i$. To obtain a contradiction, we will show that $\alpha^{n}_i(p;\rl,\mu)$ converges to 0 for at least $N-q+1$ voters.
  
  Let $i\in\{q,\dots,N\}$ and consider the following cases depending on the limit behavior of sequence $\{\mathbb P_i^n(q-1\mid\ell)\}_{n=1}^\infty$.

  \textit{(i)} The sequence is bounded away from 0. In this case, $\mathbb P_i^n(q-1)$ is bounded away from 0, and therefore \eqref{difference1} determines $i$'s voting behavior for sufficiently large $n$. Dividing \eqref{difference1} by $\mathbb P_i^n(q-1\mid\ell)$, observe that the ratio $\mathbb P_i^n(q-1\mid h)/\mathbb P_i^n(q-1\mid\ell)$ is bounded. Using $\tau^{\ell,n}_i(\rl)\to 1$ and $\tau^{h,n}_i(\rl)\to 0$, we can see that $\alpha_i^{n}(p;\rl,\mu)=0$ since $u^\ell_i(p)-u^\ell_i(0)<0$ when $p>y^\ell_i\geq y^\ell_q$ (recall that the voters are ordered with respect to their ideal policies).

  \textit{(ii)} The sequence is eventually constant at 0. In this case, $\mathbb P_i^n(q-1)=0$ for sufficiently large $n$ and therefore \eqref{difference2} determines $i$'s voting behavior. Again, $\alpha_i^{n}(p;\rl,\mu)=0$ for the same reason as above.

  \textit{(iii)} The sequence converges to 0. In this case, $\lim_{n\to\infty}\sum_{j\neq i}\alpha_j^{n}(p;\rl,\mu)\geq q$ and \eqref{difference1} determines $i$'s voting behavior. By Lemma \ref{ranking:greater}, we have $\mathbb P_i^n(q-1\mid \ell)\geq\mathbb P_i^n(q-1\mid h)$ for sufficiently large $n$, which again implies $\alpha_i^{n}(p;\rl,\mu)=0$.

  \textit{(b)} Now, suppose $\{\sum_{r\geq q}\mathbb P^n(r\mid h)\}_{n=1}^\infty$ does not converge to 1. By Lemma \ref{bound:high}, there are at least $(N-q+1)$ players $i$ for which there exists $\varepsilon_i>0$ such that for each $m$ there is $n\geq$ with the property that $\alpha_i^{n}(p;\rh,\mu)<1-\varepsilon_i$. To obtain a contradiction, we will show that $\alpha^{n}_i(p;\rh,\mu)$ converges to 1 for at least $q$ voters.

  Let $i\in\{1,\dots,q\}$ and consider the following cases depending on the limit behavior of sequence $\{\mathbb P_i^n(q-1\mid h)\}_{n=1}^\infty$.
  
  \textit{(i)} The sequence is bounded away from 0. In this case, $\mathbb P_i^n(q-1)$ is bounded away from 0, and therefore \eqref{difference1} determines $i$'s voting behavior for sufficiently large $n$. Dividing \eqref{difference1} by $\mathbb P_i^n(q-1\mid h)$, observe that the ratio $\mathbb P_i^n(q-1\mid \ell)/\mathbb P_i^n(q-1\mid h)$ is bounded. Using $\tau^{\ell,n}_i(\rh)\to 0$ and $\tau^{h,n}_i(\rh)\to 1$, we can see that $\alpha_i^{\rh,n}=1$ since $u^h_i(p)-u^h_i(0)>0$ when $p<y^h_i\leq y^h_q$.

  \textit{(ii)} The sequence is eventually constant at 0. In this case, $\mathbb P_i^n(q-1)=0$ for sufficiently large $n$ and therefore \eqref{difference2} determines $i$'s voting behavior. Again, $\alpha_i^{n}(p;\rh,\mu)=1$ for the same reason as above.

  \textit{(iii)} The sequence converges to 0. In this case, $\lim_{n\to\infty}\sum_{j\neq i}\alpha_j^{n}(p;\rh,\mu)\leq q-2$ and \eqref{difference1} determines $i$'s voting behavior. By Lemma \ref{ranking:lower}, we have $\mathbb P_i^n(q-1\mid \ell)\leq\mathbb P_i^n(q-1\mid h)$ for sufficiently large $n$, which again implies $\alpha_i^{n}(p;\rh,\mu)=1$.
\end{proof}

Lemmas \ref{benchmark:belowabove} and \ref{benchmark:between} mean that the $q$-majority voting fully aggregates the private information of voters in the sense that the limit probability that policy $p$ is accepted in state $\omega$ when signals are private is the same as the limit probability that policy $p$ is accepted in state $\omega$ when signals are public. For example, the probability that each voter receives signal $\rh$ converges to 1 in state $h$. Since every policy $p\in(y^\ell_q,y^h_q)$ is strictly preferred to the status quo in state $h$ by at least $q$ voters (namely, voters in $\{1,\dots,q\}$), the probability that $p$ is accepted in state $h$ converges to 1 when the signals are public and thus no additional information is generated by conditioning on being pivotal. This result shows that full information equivalence holds for any $q$-majority rule in a finite committee without communication.

\begin{proof}[Proof of Benchmark \ref{bench:take}]
  By Lemma \ref{benchmark:belowabove}, for each $n\geq1$ the agenda-setter's equilibrium expected payoff equals $V_A^{T,n}(\mu^0)=\max\left\{y^\ell_q,\max_{p\in(y^\ell_q,y^h_q]}\sum_{\omega\in\{\ell,h\}}\mu^0(\omega)\sum_{r\geq q}\mathbb P^n(r\mid \omega)p\right\}$, where the maximum over $p\in(y^\ell_q,y^h_q]$ exists whenever the corresponding supremum is strictly greater than $y^\ell_q$ by requiring that voters break ties in favor of acceptance. By Lemma \ref{benchmark:between} and the continuity of maximum, sequence $\{V_A^{T,n}(\mu^0)\}_{n=1}^\infty$ converges to $\max\{y^\ell_q,\mu^0y^h_q\}$.
\end{proof}

\section{Proofs}\label{sec-proofs}

Consider the difference in the expected payoffs of voter $i$ from voting to accept and voting to reject proposal $p$ given signal $s_i$ and the strategy profile $\sigma_{-i}$ of other players. If the probability that $i$ is pivotal is positive, $\mathbb P_i(q-1)>0$ (where the dependence on proposal and voting strategies is implicit), then the difference in the expected payoffs, denoted by $\mathbb EU_i(p;s_i,\mu,\alpha_{-i})$ is given by \eqref{difference3}. This expression captures both the pivotal and signaling incentives of voter $i$ with signal $s_i$ deciding how to vote for proposal $p$. In particular, even when the probability that voter $i$ is pivotal equals 0, the probability that some other voters vote to accept $p$ may be greater than 0. In this case, voter $i$ must consider the difference in continuation payoff from voting to accept and voting to reject proposal $p$ arising from the possible difference in public belief caused by the action of voter $i$.

If the probability that the $(q-1)$ or fewer voters other than $i$ vote to accept policy $p$, then the difference in \eqref{difference3} equals 0 and voter $i$ votes sincerely, that is, voter $i$ votes based on the difference in the expected payoffs written as follows:
\begin{equation}\label{difference4}
  \mathbb EU_i(p;s_i,\mu) = \sum_{\omega\in\{\ell,h\}}\mu(\omega)\tau_i^\omega(s_i)(u_i^\omega(p)-u_i^\omega(0)).
\end{equation}

The following result establishes the limit (as the signals become perfectly precise) of a lower bound of the support of policies that can be implemented in any equilibrium irrespective of state. This limit is given by policy $y^\ell_q$, implying that for any policy $p<y^\ell_q$, the probability that $p$ is implemented in any equilibrium converges to 0 in both states as the signals become perfectly precise.

\begin{lem}\label{lem:low}
  Fix $\delta<1$ and take a sequence $\{(\tau^n_i)_i\}$ such that $\tau^n_i$ converges to 1 for each $i\in\mathcal N$ and let $\{\sigma^n\}$ be a sequence of equilibrium strategy profiles. There exists a sequence of policies $\{p^n\}$ converging to $y^\ell_q$ such that the sequence $\{\sum_{r\geq q}\mathbb P^n(r\mid\omega)\}_{n=1}^\infty$ converges to 1 for both $\omega\in\{\ell,h\}$.
\end{lem}
\begin{proof}[Proof of Lemma \ref{lem:low}]
  By the monotonicity of voting strategies, we only need to prove the statement for state $\ell$. For any policy $p\in (0,y^\ell_N/2)$, each voter $i\in\mathcal N$ in each state $\omega\in\{\ell,h\}$ strictly prefers $p$ to any $p'<p$ and to the status-quo (recall that voters are ordered with respect to their ideal policies and the ideal policy of each voter in state $\ell$ is smaller than the ideal policy in state $h$), implying that $\alpha_i^{n}(p;s_i,\mu)=1$ for each $i\in\mathcal N$, $s_i\in\{\rl,\rh\}$, $\mu\in\Delta(\{\ell,h\})$, and $n\geq1$. Let $p^*=\inf_{\mu\in\Delta(\Omega)}\left\{\sup_{p\leq y^\ell_q}\left\{p\mid \lim_{n\to\infty}\sum_{r\geq q}\mathbb P^n(r\mid\ell)=1\right\}\right\}$, which is well-defined and satisfies $p^*>0$. For instance, for $p=y^\ell_N/4$ we have $\mathbb P^n(N\mid\ell)=1$ for each $n\geq 1$ as argued above.

  If $p^*=y^\ell_q$, then the proof is complete, so suppose $p^*<y^\ell_q$ and consider $p=p^*+\varepsilon$ for a small $\varepsilon>0$. Then, there exists a belief $\mu$ such that $\{\sum_{r\geq q}\mathbb P^n(r\mid\ell)\}_{n=1}^\infty$ does not converge to 1. By Lemma \ref{bound:high}, there exist $(N-q+1)$ voters $i$ such that $\alpha^{n}_{i}(p;\rl,\mu)$ is bounded away from 1. But any voter $i\in\{1,\dots,q\}$ prefers policy $p$ to any policy $p'\in[p^*,p)$ after a period of delay when $\varepsilon$ is sufficiently small (recall that $\delta<1$ is fixed), which implies that $\alpha_i^{n}(p;\rl,\mu)$ converges to 1 for each $i\in\mathcal\{1,\dots,q\}$. But then $\{\sum_{r\geq q}\mathbb P^n(r\mid\ell)\}_{n=1}^\infty$ converges to 1, which is a contradiction.
\end{proof}

The following result uses Lemma \ref{lem:low} to establish the limit of the upper bound of the support of policies that can be implemented in any equilibrium in state $\ell$. The analogous result for state $h$ requires more work and is given in Lemma \ref{lem:high2}. 

\begin{lem}\label{lem:high}
  Take a sequence $\{(\tau^n_i)_i\}$ such that $\tau^n_i$ converges to 1 for each $i\in\mathcal N$ and let $\{\sigma^n\}$ be a sequence of equilibrium strategy profiles. For each $p\in(y^\ell_q,M]$, sequence $\{\sum_{r\geq q}\mathbb P^n(r\mid\ell)\}_{n=1}^\infty$ converges to 0.
\end{lem}
\begin{proof}[Proof of Lemma \ref{lem:high}]
  To establish a contradiction, suppose that $\{\sum_{r\geq q}\mathbb P^n(r\mid\ell)\}_{n=1}^\infty$ is bounded away from 0, so Lemma \ref{bound:low} implies that for at least $q$ voters $\{\alpha^{n}_i(p;\rl,\mu)\}$ is bounded away from 0. We will obtain a contradiction by showing that $\alpha^{n}_i(p;\rl,\mu)$ converges to 0 for at least $N-q+1$ voters.
  
  Fix voter $i\in\{q,\dots,N\}$ and consider the following cases depending on the limit behavior of sequence $\{\mathbb P_i^n(q-1\mid \ell)\}_{n=1}^\infty$.
  
  \textit{(i)} The sequence is eventually constant at 0. In this case, $\alpha^{n}_j(p;\rl,\mu)=1$ for at least $q$ voters other than $i$, which together with monotonicity of voting strategies implies $\sum_{r\leq q-1}\mathbb P_i^n(r\mid \omega)=0$ for both $\omega\in\{\ell,h\}$. Therefore, \eqref{difference4} determines $i$'s voting behavior. Since $i$ strictly prefers the status quo to any $p>y^\ell_q\geq y^\ell_i$, we have $\alpha_i^{n}(p;\rl,\mu)=0$.

  \textit{(ii)} The sequence converges to 0. In this case, $\lim_{n\to\infty}\sum_{j\neq i}\alpha_j^{n}(p;\rl,\mu) \geq q$ and \eqref{difference3} determines $i$'s voting behavior. By monotonicity, $\lim_{n\to\infty}\sum_{j\neq i}\alpha_j^{n}(p;\rh,\mu) \geq q$ and the mode of the Poisson binomial distribution with parameter vector $\alpha_{-i}^{n}(p;\rh,\mu)$ is at least $(q-1)$ for sufficiently large $n$. Therefore, we have $\mathbb P_i^n(q-1\mid \ell)\geq\mathbb P_i^n(q-1\mid h)$ by Lemma \ref{ranking:greater} and $\mathbb P_i^n(r\mid h)<\mathbb P_i^n(q-1\mid h)$ for all $r<q-1$ by Lemma \ref{ref:poisson}, both for sufficiently large $n$.

  Along a subsequence where $\mathbb P_i^n(q-1\mid \ell)>0$, we can divide \eqref{difference3} by $\mathbb P_i^n(q-1\mid \ell)$ for each $n$ and consider its limit. Given signal $s_i=\rl$, the terms related to state $h$ vanish and $\alpha_i^{n}(p;\rl,\mu)=0$ since the pivotal and signaling incentives of voter $i$ are aligned in state $\ell$ and $u_i^\ell(p)-u_i^\ell(0)<0$. To see this, note that on the one hand, voter $i$ strictly prefers $p'\in[y^\ell_q,p]$ after a period of delay to policy $p>y^\ell_q\geq y^\ell_i$. On the other hand, voter $i$ prefers a first-order stochastically dominated distribution of implemented policies since $i$'s payoff is decreasing on $[y^\ell_q,M]$.

  \textit{(iii)} The sequence is bounded away from 0. The argument above continues to hold.
\end{proof}

The following result is similar to Lemma \ref{lem:low} and refines the limit of the lower bound of policies that can be implemented in any equilibrium when the state is $h$. This revised limit holds only in case the conflict in preferences of voter $q$ between the states is sufficiently high.

\begin{lem}\label{lem:low2}
  Let $y^\ell_q<y^h_q-y^\ell_q$, take a sequence $\{(\tau^n_i)_{i\in\mathcal N}\}_{n=1}^\infty$ such that $\tau^n_i$ converges to 1 for each $i\in\mathcal N$, and let $\{\sigma^n\}_{n=1}^\infty$ be a sequence of equilibrium strategy profiles. There exists a sequence of policies $\{p^n\}_{n=1}^\infty$ converging to $(y^h_q-y^\ell_q)$ such that the sequence $\{\sum_{r\geq q}\mathbb P^n(r\mid h)\}_{n=1}^\infty$ converges to 1.
\end{lem}
\begin{proof}[Proof of Lemma \ref{lem:low2}]
  The idea is similar to the proof of Lemma \ref{lem:low}. Consider policy $p=y^h_q/2$, which is the ideal policy of voter $q$ in state $h$. We want to show that the probability that $p$ is accepted in state $h$ converges to 1. Suppose it does not. Then, Lemma \ref{bound:high} implies that there are at least $(N-q+1)$ players $i$ for which there exists $\varepsilon_i>0$ such that for each $m$ there is $n\geq m$ with the property that $\alpha_i^{n}(p;\rh,\mu)<1-\varepsilon_i$. We will obtain a contradiction by showing that $\alpha^{n}_i(p;\rh,\mu)$ converges to 1 for at least $q$ voters.
  
  Let $i\in\{1,\dots,q\}$ and consider a sequence $\{\sum_{r\leq q-1}\mathbb P_i^n(r\mid h)\}_{n=1}^\infty$, which is bounded away from 0. Dividing \eqref{difference3} by $\sum_{r\leq q-1}\mathbb P_i^n(r\mid h)$ we can notice that the terms corresponding to state $\ell$ vanish as signals become perfectly precise. But the pivotal and signaling incentives of voter $i$ in state $h$ are aligned when voting on policy $p=y^h_q/2\leq y^h_i/2$, and therefore $\alpha^{n}_j(p;\rh,\mu)=1$.

  Now, define $p^*=\inf_{\mu\in\Delta(\Omega)}\left\{\sup_{p\leq y^h_q-y^\ell_q}\left\{p\mid \lim_{n\to\infty}\sum_{r\geq q}\mathbb P^n(r\mid h)=1\right\}\right\}$, suppose $p^*<y^h_q-y^\ell_q$ and consider policy $p=p^*+\varepsilon$ for a small $\varepsilon>0$. Lemma \ref{lem:low} and assumption \eqref{coase-dynamics} imply that for each $\varepsilon'>0$, the policy that is eventually implemented is contained in $\varepsilon'$-neighborhood of set $\{y^\ell_q,p^*\}$ with probability converging to 1. Therefore each voter $j\in\{1,\dots,q\}$ in state $h$ strictly prefers to accept $p$. The argument is completed by showing that only preferences in state $h$ matter for voter $j$ with signal $\rh$ when the probability that $p$ is accepted in state $h$ does not converge to 1. This can be done analogously to the proof that $y^h_q/2$ is accepted in state $h$ with probability converging to 1.
\end{proof}

Lemma \ref{lem:low2} allows us to establish the analog of Lemma \ref{lem:high} for state $h$.

\begin{lem}\label{lem:high2}
  Take a sequence $\{(\tau^n_i)_{i\in\mathcal N}\}_{n=1}^\infty$ such that $\tau^n_i$ converges to 1 for each $i\in\mathcal N$ and let $\{\sigma^n\}_{n=1}^\infty$ be a sequence of equilibrium strategy profiles. For each $p\in(y^h_q,M]$, sequence $\{\sum_{r\geq q}\mathbb P^n(r\mid h)\}_{n=1}^\infty$ converges to 0.
\end{lem}
\begin{proof}[Proof of Lemma \ref{lem:high2}]
  To establish a contradiction, suppose that $\{\sum_{r\geq q}\mathbb P^n(r\mid h)\}_{n=1}^\infty$ is bounded away from 0, so Lemma \ref{bound:low} implies that for at least $q$ voters $\{\alpha^{n}_i(p;\rh,\mu)\}$ is bounded away from 0. We will show that $\alpha^{n}_i(p;\rh,\mu)$ converges to 0 for at least $N-q+1$ voters.

  Fix voter $i\in\{q,\dots,N\}$ and consider the following cases depending on the limit behavior of sequence $\{\mathbb P_i^n(q-1\mid h)\}_{n=1}^\infty$.

  \textit{(i)} The sequence is eventually constant at 0. In this case, $\alpha^{n}_j(p;\rl,\mu)=1$ for at least $q$ other players and $\sum_{r\leq q-2}\mathbb P_i^n(r\mid \omega)=0$ for both $\omega\in\{\ell,h\}$. Using \eqref{difference4}, sincere voting by voter $i$ implies $\alpha_i^{n}(p;\rh,\mu)=0$ by the choice of voter $i$ and proposal $p$.

  \textit{(ii)} The sequence converges to 0 or bounded away from 0. In this case, voter $i$'s strategy is determined by \eqref{difference3}. Notice that in both states voter $i$ strictly prefers $p'\in[y^h_q,p]$ after a period of delay to policy $p$, and prefers a first-order stochastically dominated distribution of implemented policies since her payoff is decreasing on $[y^h_q,M]$. Therefore, the signaling and pivotal incentives are aligned in both states and we must have $\alpha^{\rh,n}_i=0$.
\end{proof}

Finally, Lemmas \ref{lem:low}-\ref{lem:high2} imply the following lower and upper limits on the agenda-setter's expected payoffs.
\begin{lem}\label{lem:bounds}
  Take a sequence $\{(\tau^n_i)_{i\in\mathcal N}\}_{n=1}^\infty$ such that $\tau^n_i$ converges to 1 for each $i\in\mathcal N$ and let $\{\sigma^n\}_{n=1}^\infty$ be a sequence of equilibrium strategy profiles. Then:
  \begin{enumerate}[(a)]
    \item For each $\varepsilon>0$, there exists $m$ such that the agenda-setter's expected payoff $V_{A,n}$ is at most $\mu^0(h)y^h_q+\mu^0(\ell)y^\ell_q+\varepsilon$ for all $n\geq m$.
    \item For each $\varepsilon>0$, there exists $m$ such that the agenda-setter's expected payoff $V_{A,n}$ is at least $(y^\ell_q-\varepsilon)$ for all $n\geq m$.
  \end{enumerate}
\end{lem}

Lemmas \ref{lem:low} and \ref{lem:high} have an immediate implication for the agenda-setter's behavior when she becomes sufficiently pessimistic about the state. The following result shows that the agenda-setter does not screen when the belief that the state is $h$ is below a certain threshold, meaning that the agenda-setter makes a proposal accepted with a probability approaching 1 in both states.

\begin{lem}\label{lem:screening}
  Take a sequence $\{(\tau^n_i)_{i\in\mathcal N}\}_{n=1}^\infty$ such that $\tau^n_i$ converges to 1 for each $i\in\mathcal N$ and let $\{\sigma^n\}_{n=1}^\infty$ be a sequence of strategy profiles. There exists a sequence of policies $\{p^n\}_{n=1}^\infty$ and belief thresholds $\{\mu^n\}_{n=1}^\infty$, such that the agenda-setter proposes $p^n$ if the public belief is below $\mu^n$. Moreover, $\{p^n\}_{n=1}^\infty$ converges to $y^\ell_q$ and the corresponding sequence of acceptance probabilities converges to 1.
\end{lem}

\begin{proof}[Proof of Lemma \ref{lem:screening}]
  We know from Lemma \ref{lem:low} that for a sufficiently large $n$ the agenda-setter can guarantee the expected payoff $V^n$ by proposing some policy $p^n$, where $\{p^n\}_{n=1}^\infty$ and $\{V^n\}_{n=1}^\infty$ both converge to $y^\ell_q$. In particular, the probability that $p^n$ is accepted converges to 1 in both states $\omega\in\{\ell,h\}$. At the same time, we know from Lemma \ref{lem:high} the agenda-setter's expected payoff in state $\ell$ is bounded above by $(y^\ell_q+\varepsilon^n)$ where $\{\varepsilon^n\}_{n=1}^\infty$ converges to 0.

  Suppose that there exists a sequence of screening equilibria with initial proposals $\{\tilde p^n\}_{n=1}^\infty$, meaning that the probability that $\tilde p^n$ is accepted is at most $(1-\varepsilon)$ for some $\varepsilon>0$. It follows that the agenda-setter's expected payoffs along this sequence are bounded above by sequence $\{\mu(h) M + \delta \mu(\ell)(y^\ell_q+\varepsilon^n)\}_{n=1}^\infty$. But, the agenda-setter will strictly prefer to make proposal $p^n$ when the public belief that the state is $h$ is below $\mu^n=\frac{V^n-\delta(y^\ell_q+\varepsilon')}{M-\delta(y^\ell_q+\varepsilon')}$, which contradicts the existence of such sequence of screening equilibria.
\end{proof}

\subsection{Equal signal precision}

In this section, we shown that the upper bound in Lemma \ref{lem:bounds} is tight when the private signals of players have equal precision.

\begin{proof}[Proof of Theorem \ref{theorem-equal}]
  The strategies of voters $(q-1)$ and $q$ are presented in Section \ref{sec-ideas}. Let $V_i(\omega,\mu)$ be voter $i$'s expected payoff in state $\omega$ given belief $\mu$. It follows from Lemma \ref{lem:screening} that $V_i(\omega,\mu)$ converges to $u^\omega_i(y^\ell_q)$ when $\mu(h)$ is sufficiently close to 0. Let $\tilde p_q$ be the largest policy $p$ such that the following inequalities are satisfied for $i\in\{q-1,q\}$:
  \begin{align*}
    0 \geq\; & \mu(\ell)\tau^2\big[\delta V_{i}(\ell,\mu)-\delta V_i(\ell,\tilde \mu)\big] + \mu(\ell)\tau(1-\tau)\big[u_{i}^\ell(p)-(1-\delta)u^\ell_{i}(0)-\delta V_{i}(\ell,\mu)\big] \nonumber\\
    +\; & \mu(h)(1-\tau)^2\big[\delta V_{i}(h,\mu) - \delta V_i(h,\tilde \mu)\big] + \mu(h)(1-\tau)\tau\big[u^h_{i}(p) - (1-\delta)u^h_{i}(0) - \delta V_{i}(h,\mu)\big], \\
    0 \geq \; & \mu(\ell)(1-\tau)^2\big[\delta V_i(\ell,\tilde \mu) - \delta V_{i}(\ell,\mu)\big] + \mu(\ell)(1-\tau)\tau\big[(1-\delta)u^\ell_{i}(0)+\delta V_{i}(\ell,\mu) - u_{i}^\ell(p)\big] \nonumber\\
    +\; & \mu(h)\tau(1-\tau)\big[\delta V_i(h,\tilde \mu) - \delta V_{i}(h,\mu)\big] + \mu(h)\tau^2\big[(1-\delta)u^h_{i}(0)+\delta V_{i}(h,\mu) - u^h_{i}(p)\big],
  \end{align*}
  By construction, voters in $\{1,\dots,q-2\}$ vote to accept $\tilde p_q$ and voters in $\{q+1,\dots,N\}$ vote to reject it. Therefore, by the definition of $\tilde p_q$, voters $(q-1)$ and $q$ do not have profitable deviations. To complete the proof, we need to show that voters in $\{1,\dots,q-2\}$ and $\{q+1,\dots,N\}$ do not have an incentive to deviate.

  First, notice that $\lim_{\tau\to 1}\sup_{\mu,\mu'}\lvert \tilde p_q(\mu)-\tilde p_q(\mu')\rvert=0$. Second, notice that for each voter $i\in\mathcal N$, given the tentative strategy profile, the probability that $(q-1)$ or fewer other voters vote to accept $\tilde p_q$ is strictly positive, meaning that voters use \eqref{difference3} to evaluate the deviations. Consider voter $i\in\{1,\dots,q-2\}$. If voter $i$ deviates and votes to reject $\tilde p_q$, there are three possibilities:

  \textit{(i)} Voters $(q-1)$ and $q$ receive high signals, and the posterior public belief that the state is $h$ is close to 1. A policy proposal close to $\tilde p_q$ is made and voter $i$ believes for each $s_i\in\{\rl,\rh\}$ the proposal is accepted with probability close to 1.

  \textit{(ii)} Voters $(q-1)$ and $q$ receive conflicting signals, and the public belief remains the same. A policy proposal $\tilde p_q$ is made again.

  \textit{(iii)} Voters $(q-1)$ and $q$ receive low signals, and the posterior public belief that the state is $h$ is close to 0. A policy proposal close to $y^\ell_q$ is made, and voter $i$ believes for each $s_i\in\{\rl,\rh\}$ the proposal is accepted with probability close to 1.

  Since voter $i$ is almost indifferent between policy $\tilde p_q$ and the revised policy and strictly prefers both policies to the status quo, the deviation is not profitable in cases \textit{(i)} and \textit{(ii)}. In case \textit{(iii)}, the action of voter $i$ does not affect the distribution of outcomes.

  Next, consider voter $i\in\{q+1,\dots,N\}$. The action of voter $i$ does not affect the public belief if at least $(n-q+2)$ other voters reject $\tilde p_q$. And when $(q-1)$ other voters vote to accept $\tilde p_q$, voter $i$ strictly prefers delay to policy $\tilde p_q$ since $\tilde p_q$ is close to $y^\ell_q$ and therefore greater than $y^\ell_i$.
\end{proof}

\subsection{Unequal signal precision}

\begin{lem}\label{lem:upper:unequal}
  Take a sequence $\{(\tau^n_i)_{i\in\mathcal N}\}_{n=1}^\infty$ such that $\{\tau^n_i\}_{n=1}^\infty$ converges to 1 for each $i\in\mathcal N$ and let $\{\sigma^n\}_{n=1}^\infty$ be a sequence of strategy profiles. Moreover, suppose that $\{\frac{1-\tau_i^n}{\prod_{j\neq i}(1-\tau_j^n)}\}_{n=1}^\infty$ converges to 0 for some $i\in\mathcal N$. For each $p\in(\min\{y^h_q,\max\{y^\ell_q,y^h_i-y^\ell_q\}\},M]$, sequence $\{\sum_{r\geq q}\mathbb P^n(r\mid h)\}_{n=1}^\infty$ converges to 0.
\end{lem}

\begin{proof}[Proof of Lemma \ref{lem:upper:unequal}]
  Lemma \ref{lem:high} implies that the result holds for $p\in[y^h_q,M]$, so it is sufficient to consider $p\in(y^h_i-y^\ell_q,M]$ where $(y^h_i-y^\ell_q)\in(y^\ell_q,y^h_q)$.

  Suppose $p>y^h_i-y^\ell_q$ is accepted in state $h$ with probability bounded away from 0. Then, there exist at least $q$ voters $i$ such that $\alpha_i^{\rh,\mu}(p)$ bounded away from 0. Note that players in $\{q+1,\dots,N\}$ strictly prefer to reject $p$ after receiving signal $\rh$ when the signals are sufficiently precise given the assumptions that $p>y^h_i-y^\ell_q$ and $i<q$. Therefore, $\alpha_k^{\rh,\mu}(p)$ is bounded away from 0 for each $j\in\{1,\dots,q\}$. Consider two cases based on the behavior of sequence $\{\alpha_{i}^{n}(p;\rh,\mu)-\alpha_{i}^{n}(p;\rl,\mu)\}_{n=1}^\infty$.
  
  Suppose $\{\alpha_{i}^{n}(p;\rh,\mu)-\alpha_{i}^{n}(p;\rl,\mu)\}_{n=1}^\infty$ converges to 0. In that case, we must have $\alpha_{i}^{n}(p;\rl,\mu)$ bounded away from 0, which is a contradiction since voter $i$ with signal $\rl$ believes that the state is $\ell$ and: (a) believes that any proposal $p>y^\ell_q$ will be rejected with probability approaching 1; and (b) in state $\ell$ prefers $y^\ell_q$ to $p>y^h_i-y^\ell_q$.

  If $\{\alpha_{i}^{n}(p;\rh,\mu)-\alpha_{i}^{n}(p;\rl,\mu)\}_{n=1}^\infty$ is bounded away from 0, then voter $i$ with signal $\rh$ has a profitable deviation when the signals are sufficiently precise. Such voter believes that the state is $h$ and: (a) rejection leads to a revised belief that puts less weight on state $h$; and (b) in state $h$ prefers any policy $p'\leq y^h_i-y^\ell_q$ to $p>y^h_i-y^\ell_q$.
\end{proof}

\begin{proof}[Proof of Theorem \ref{theorem-unequal}]
  We need to construct a sequence of equilibria for which the agenda-setter's expected payoff converges to the upper bound identified in Lemma \ref{lem:upper:unequal}, that is, $\mu(h)\max\{y^\ell_q,y^h_q-y^\ell_q\}+\mu(\ell)y^\ell_q$. Construct a sequence of equilibrium profiles as follows. Every player $j\neq i$ uses an uninformative voting strategy (to be specified later), so the agenda-setter makes inferences based only on the actions of voter $i$, and voter $i$ does not receive any equilibrium information. From here, the construction mirrors the construction of the equilibrium path in bilateral bargaining over price. However, the single-peaked preferences imply a different limiting behavior of the initial proposal and, therefore, the agenda-setter's expected payoff.
 
  For each policy $p$, signal $s_i$, and belief $\mu$, let $U_i(p;s_i,\mu) = \mathbb E[u^\omega_i(p)\mid s_i,\mu]$ be the expected utility of voter $i$ from policy $p$. Then, let policy $p^{s_i,\mu}_i$ be the maximum policy that voter $i$ with signal $s_i$ and prior belief $\mu$ weakly prefers to the status-quo, i.e., $p^{s_i,\mu}_i = \max \{p\in[0,M] \mid U_i(p;s_i,\mu) \geq U_i(0;s_i,\mu)\}$. For any belief $\mu$, as the signal precision $\tau$ converges to one, policy $p^{\rl,\mu}_i$ converges to $y^\ell_i$ and policy $p^{\rh,\mu}_i$ converges to $y^h_i$. These are the highest policies that voter $i$ with corresponding signals can accept in any equilibrium.

  It is straightforward to show that under the skimming assumption \eqref{coase-dynamics}, policy $p_i^{\rl,\mu}$ must be accepted with certainty when the public belief about the state $h$ is sufficiently low. Therefore, when $\mu(h)\leq p_i^{\rl,\mu}/p_i^{\rh,\mu}$, the agenda-setter proposes policy $p_i^{\rl,\mu}$ which is accepted with certainty.

  Now, construct a sequence of policies $\{p_t\}_{t=1}^\infty$ and a sequence of belief cutoffs $\{\m_t\}_{t=1}^\infty$ as follows. Denote $p_1=p^{\rl,\mu}_i$ and $\m_1=p^{\rl,\mu}_i/p^{\rh,\mu}_i$. Then, define $p_2^\mu$ and $\m_2$ as:
  \begin{align}
    p_2^\mu & = \max\{p\in[0,M]\mid U_i(p;\rh,\mu) \geq  (1-\delta)U_i(0;\rh,\mu) + \delta U_i(p_1;\rh,\mu)\}, \label{def:p2} \\
    \m_2 & =\max\Bigg\{m\geq \m_1\mid mU_A(p_1^m) + (1-m)\delta U_A(p_1) \nonumber\\
    & \hspace{1in}\left. \geq \frac{m-\m_1}{1-\m_1}U_A(p_2^m,m)+\left(1-\frac{m-\m_1}{1-\m_1}\right)\delta U_A(p_1)\right\}. \label{def:m2}
  \end{align}

  Recall that $G^{\omega,\mu}$ denotes the distribution over pairs $(t,p)$ induced by strategies of players in state $\omega$ given the public belief $\mu$, where the dependence on strategies is implicit. When the support of $G^{\omega,\mu}$ is discrete, which is the case in this section, then $G^{\omega,\mu}(t,p)$ is the probability that policy $p$ is implemented in $t$ periods. We can write the expected continuation payoffs given belief $\mu$ and signal $s_i=\rh$ for a continuation path implied by the rejection of $p$ leading to belief $\mu'$ as follows:
  \begin{align}
    V_i(p;\rh,\mu) = \sum_{\omega\in\{\ell,h\}}\mu(\omega\mid \rh)\sum_{(t',p')}\delta^{t'}u_i^\omega(p')G^{\omega,\mu'}(t', p'), \label{def:vi}\\ 
    V_A(p;\mu) = \sum_{\omega\in\{\ell,h\}}\mu(\omega)\sum_{(t',p')}\delta^{t'}u_A(p')G^{\omega,\mu'}(t', p'), \label{def:va}
  \end{align}
  where the second summation is over the support of $G^{\omega,\mu}$. These expressions were already used in the definitions \eqref{def:p2} and \eqref{def:m2} for a degenerate distribution $G^{\omega,\mu}$ with support $\{(0,p_1)\}$. We can use \eqref{def:vi} and \eqref{def:va} to define $p_t$ and $\m_t$ recursively for $t\geq 3$ as follows:
  \begin{align}
    p_t^\mu & = \max\{p\in[0,M]\mid U_i(p;\rh,\mu) \geq  (1-\delta)U_i(0;\rh,\mu) + \delta V_i(p;\rh,\mu)\}, \label{def:pt} \\
    \m_t & =\max\Bigg\{m\geq \m_{t-1}\mid mU_A(p_{t-1}^m) + (1-m)\delta V_A(p_{t-1};\mu) \nonumber\\
    & \hspace{1in}\left. \geq \frac{m-\m_{t-1}}{1-\m_{t-1}}U_A(p_t^m,m)+\left(1-\frac{m-\m_{t-1}}{1-\m_{t-1}}\right)\delta V_A(p_t^m;m)\right\}. \label{def:mt}
  \end{align}
  All there remains to do is specify players' strategies following a rejection of $p^\mu_t$, which will provide us with the distribution $G^{\omega,\mu}$. When the public belief $\mu(h)\in[0,\m_1]$, the agenda-setter proposes $p_1$ and voter $i$ follows strategy $\bm1(p\leq p_1)$ for both signals. For $t\geq2$, when the public belief $\mu(h)\in(\m_{t-1},\m_t]$, the agenda-setter proposes $p_t$ and voter $i$ uses the following strategy:
  \begin{equation}
    \alpha_{i}(p;\rl,\mu) = \bm1(p\leq p_1); \qquad \alpha_{i}(p;\rh,\mu) =
    \begin{cases}
      1 & \text{if } p\leq p_{t-1}, \\
      \varphi^{\mu(h)}_t & \text{if } p_{t-1}<p\leq p_t, \\
      0 & \text{if } p>p_{t},
    \end{cases}
  \end{equation}
  where $\varphi_t^{\mu(h)}$ is such that the posterior belief that the state is $h$ is $m{t-1}$ when $p$ is rejected, that is, $\varphi_t^{\mu(h)}$ solves $\m_{t-1}=\frac{\mu(h)(1-\tau_i\varphi_{t}^{\mu(h)})}{\mu(h)(1-\tau_i\varphi_{t}^{\mu(h)})+\mu(\ell)(1-(1-\tau_i)\varphi_{t}^{\mu(h)})}$. Given these strategies, the agenda-setter follows a decreasing sequence of proposals $\{p_T, \dots, p_1\}$ where $T$ is determines by the prior belief $\mu^0(h)\in(\m_T,\m_{T+1}]$.

  Turning to voters $j\neq i$, suppose that for each signal $s_j\in\{\rl,\rh\}$ and belief $\mu\in\Delta(\{\ell,h\})$: each voter $j\neq i$ in $\{1,\dots,q\}$ votes to accept every proposal $p\leq y^{\rh,\mu}_q$ and reject every proposal $p>y^{\rh,\mu}_q$, and each voter $j$ in $\{q+1,\dots,N\}$ votes to accept every proposal $p\leq y^{\rl,\mu}_j$ and reject every proposal $p>y^{\rl,\mu}_j$. These strategies imply that voter $i$ is pivotal with certainty when the policy $p\in[y^{\rl,\mu}_q,y^{\rh,\mu}_q]$ is proposed given belief $\mu$.
\end{proof}

\section{Online appendix}

\subsection{Information sets and strategies}\label{appendix-strategies}

Define the following partition on the set of all possible voting profiles: $\mathrm{Acc}= \{a\in\{0,1\}^n\mid\sum_{i\in\mathcal N}a_{i}\geq q\}$ and $\mathrm{Rej} = \{a\in\{0,1\}^n\mid\sum_{i\in\mathcal N}a_{i}<q\}$. For each $t=1,2,\dots$, a history $\hbar^t$ of length $t$ can capture the game progression up until the period-$t$ proposal, can include the period-$t$ proposal or can include both the period-$t$ proposal and the voting profile. Every history $\hbar^t$ that ends with a voting profile $a_t\in\mathrm{Acc}$ is terminal. The agenda-setter moves after the state and the private signals are drawn and at histories ending with a voting profile $a_t\in\mathrm{Rej}$. Voters move at histories ending with a period-$t$ proposal.

Let $\mathcal I_A$ be the collection of information sets of the agenda-setter. The elements of $\mathcal I_A$ that include histories of length $t$ can be indexed by the history of rejected proposals $\bm p^t=(p_1,\dots,p_{t-1})$ and the correponding voting profiles $\bm a^t=(a_1,\dots,a_{t-1})$. In addition to the history of proposals and voting profiles, the period-$t$ information sets of voter $i\in\mathcal N$ are also indexed by the signal $s_i$ received by voter $i$. The collection of information sets of voter $i$ is denoted $\mathcal I_i$.

The beliefs of players about the state $\omega$ are given by a belief system $\mathcal M=\{\mu^I(\cdot)\}_{I\in\mathcal I}$, where $\mu^I(\cdot)$ is the belief of an active player in information set $I\in\mathcal I$. The common prior belief is $\mu^0(\cdot)\in\mathcal F$, where $\mathcal F=\Delta(\Omega)$ is the set of probability measures over $\Omega$. To simplify notation, I often write $\mu^0$ instead of $\mu^0(h)$ and refer to $\mu^0\in(0,1)$ as a prior belief (that the state is $h$). Similar to the prior belief $\mu^0(\cdot)$, I often write $\mu^I$ instead of $\mu^I(h)$ and refer to $\mu^I$ as a posterior belief in information set $I\in\mathcal I$.

Let $\mathcal I_A$ and $\mathcal I_i$ be the collections of information sets where the agenda-setter and player $i\in\mathcal N$ move. A (behavioral) strategy of the agenda-setter is $\pi:\mathcal I_A\to\Delta([0,M])$. In particular, for each proposal history $\bm p^t\in[0,M]^{t-1}$ with $t>1$ and voting history $\bm a^t\in\mathrm{Rej}^{t-1}$, the distribution of proposed policies in period $t$ is $\pi^{\bm p^t,\bm a^t}(\cdot)$. The distribution of the initial proposal is denoted $\pi^0(\cdot)$. A (behavioral) strategy of voter $i\in\mathcal N$ is $\alpha_i:\mathcal I_i\to[0,1]$ is the acceptance strategy of voter $i$. In particular, the probability with which voter $i$ with signal $s_i\in\{\rl,\rh\}$ accepts proposal $p_t\in[0,M]$ with $t\geq0$ after the proposal history $\bm p^t$ and (possibly empty) voting history $\bm a^t$ is $\alpha^{s_i,\bm p^t,\bm a^t}_{i,t}(p_t)$.

\subsection{Posterior beliefs and sequential rationality}
The posterior belief $\tilde \mu\in\Delta(\Omega)$ is derived from the prior belief $\mu$ given the previous proposal $p_t$ and the voting record $a_t$ as follows:
\begin{align}
  \tilde\mu^{p_t,a_t}(\omega\mid\mu) & \propto \mu(\omega)\sum_{s\in S^N}\prod_{i\in\mathcal N}\tau_i(s_i\mid\omega)(\alpha_{i,t}(p_t;s_i,\mu))^{a_{i,t}}(1-\alpha_{i,t}(p_t;s_i,\mu))^{1-a_{i,t}} \nonumber\\
  & = \mu(\omega)\prod_{i\in\mathcal N}(\alpha_{i,t}(p_t;\omega,\mu))^{a_{i,t}}(1-\alpha_{i,t}(p_t;\omega,\mu))^{1-a_{i,t}},
\end{align}
where $\tau(s_i\mid \omega)$ equals $\tau$ if both $s_i=\rl$ and $\omega=\ell$ or both $s_i=\rh$ and $\omega=h$ and equals $(1-\tau)$ otherwise, and $\alpha_{i,t}(p_t;\omega,\mu)=\sum_{s_i\in S}\tau_i(s_i\mid \omega)\alpha_{i,t}(p_t;s_i,\mu)$. In other words, $\tau_i(\cdot\mid\omega)$ is the distribution over private signals of player $i\in\mathcal N$ and $\alpha_{i,t}(p_t;\omega,\mu)$ is the probability that voter $i$ accepts proposal $p_t$ in state $\omega$. When the probability of observing a voting profile $a_t$ equals zero, I assume that the posterior belief $\tilde \mu$ equals the prior $\mu$. As elsewhere in this paper, I write $\tilde\mu^{p_t,a_t}(\mu)$ instead of $\tilde\mu^{p_t,a_t}(\omega\mid\mu)$ to denote the posterior probability that the state is $h$.

The following lemma shows that the public belief that the state is high is increasing in the action of each voter when the voting strategies are monotone. This result, combined with restriction \eqref{coase-stochastic}, is used in the proofs of Lemmas \ref{lem:low} and \ref{lem:low2}.

\begin{lem}\label{belief:increasing}
  Let $\sigma$ be a strategy profile with monotone voting strategies. Then, the posterior belief that the state is $h$ increases in the vote of each voter $i\in\mathcal N$.
  \end{lem}
  \begin{proof}[Proof of Lemma \ref{belief:increasing}]
    The posterior belief that the state is $h$ given a voting profile $a=(a_i,a_{-i})$ is given by $\tilde\mu(h\mid \mu,p,a_i,a_{-i})=\frac{\mathbb P(a_i,a_{-i}\mid h)\mu(h)}{\sum_{\omega\in\{\ell,h\}}\mathbb P(a_i,a_{-i}\mid\omega)\mu(\omega)}$, where
    \begin{align}
      \mathbb P(a_i,a_{-i}\mid \omega)&=\sum_{s\in\{\rl,\rh\}^N}\Big\{(\alpha_i^{s_i}(p))^{a_i}(1-\alpha_i(p;s_i,\mu))^{1-a_i}\prod_{j\neq i}(\alpha_j(p;s_j,\mu))^{a_j}(1-\alpha_j(p;s_j,\mu))^{1-a_j}\Big\}\tau^\omega(s) \nonumber\\
      &= \sum_{s_i}(\alpha_i(p;s_i,\mu))^{a_i}(1-\alpha_i(p;s_i,\mu))^{1-a_i}\tau_i^\omega(s_i) \nonumber\\
      &\times\sum_{s_{-i}}\Big\{\prod_{j\neq i}(\alpha_j(p;s_j,\mu))^{a_j}(1-\alpha_j(p;s_j,\mu))^{1-a_j}\Big\}\tau^\omega(s_{-i}) \label{prob:actions}
    \end{align}
    and $\tau^\omega(s)=\prod_{i\in\mathcal N}\tau_i^{\omega}(s_i)$ for all signal profiles $s$ with $\tau^\omega(s_{-i})$ defined analogously. Divide the numerator and denominator on the right-hand side of $\tilde\mu(h\mid \mu,p,a_i,a_{-i})$ by $(\alpha_i(p;s_i,\mu))^{a_i}(1-\alpha_i(p;s_i,\mu))^{1-a_i}\tau_i^\omega(s_i)$ and compare the resulting expression for $a_i=1$ with $a_i=0$. To show that the former is greater than the latter, \eqref{prob:actions} implies that it is sufficient to show that $\frac{\alpha_i(p;\rh,\mu)\tau_i+\alpha_i(p;\rl,\mu)(1-\tau_i)}{\alpha_i(p;\rh,\mu)(1-\tau_i)+\alpha_i(p;\rl,\mu)\tau_i}\geq \frac{(1-\alpha_i(p;\rh,\mu))\tau_i+(1-\alpha_i(p;\rl,\mu))(1-\tau_i)}{(1-\alpha_i(p;\rh,\mu))(1-\tau_i)+(1-\alpha_i(p;\rl,\mu))\tau_i}$, which is equivalent to $(\alpha_i(p;\rh,\mu)-\alpha_i(p;\rl,\mu))(1-(1-\tau_i)^2/\tau_i^2)\geq0$. Since $\alpha_i(p;\rh,\mu)\geq\alpha_i(p;\rl,\mu)$ by the monotonicity of voting strategies and $\tau_i\geq\frac12$, we obtain the required inequality.
\end{proof}

Sequential rationality requires that the proposal strategy $\pi(\mu)\in\Delta([0,M])$ is optimal given every belief $\mu$:
\begin{equation}
  \supp(\pi(\mu))\subseteq\argmax_{p\in[0,M]}\left\{\sum_{a\in\mathrm{Acc}}\mathbb P^{\mu,p}(a)p+\sum_{a\in\mathrm{Rej}}\mathbb P^{\mu,p}(a)\delta V_A(\tilde\mu^{p,a}(\mu))\right\},
\end{equation}
and the voting strategy of voter $i$ is optimal given every belief $\mu$, signal $s_i$, and proposal $p$:
\begin{align}
  \supp(\alpha_i(p;s_i,\mu))\subseteq\argmax_{a_i\in\{0,1\}} & \left\{\sum_{(a_i,a_{-i})\in\mathrm{Acc}}\mathbb P^{\mu,p}(a_{-i}\mid s_i)\mathbb E^{\mu,p}\left[u_i^\omega(p)\mid s_i,a_{-i}\right] \right. \nonumber\\
  & \left.+\sum_{(a_i,a_{-i})\in\mathrm{Rej}}\mathbb P^{\mu,p}(a_{-i}\mid s_i)\delta V_i(s_i,\tilde\mu^{p,a}(\mu))\right\}
\end{align}

\subsection{Ability to make revised proposals}\label{appendix-ability}

Using Theorems \ref{theorem-equal} and \ref{theorem-unequal}, we can summarize the comparison between $V_A(\mu^0)$ and $V_A^T(\mu^0)$ as follows.
\begin{cor}
  The ability to make revised proposals is:
  \begin{enumerate}
    \item Strictly valuable to the agenda-setter when one of the following conditions hold:
    \begin{enumerate}[(a)]
      \item Voters receive signals of equal precision.
      \item Voter $i\leq q$ receives an asymptotically more precise signal than other voters, $y^h_i-y^\ell_q>y^\ell_q$, and $\mu(h)<\frac{y^\ell_q}{2y^\ell_q-y^h_i+y^h_q}$.
    \end{enumerate}
    \item Strictly harmful to the agenda-setter when voter $i\leq q$ receives an asymptotically more precise signal than other voters and one of the following conditions hold:
    \begin{enumerate}[(a)]
      \item $y^h_i-y^\ell_q\leq y^\ell_q$ and $\mu(h)>\frac{y^\ell_q}{y^h_q}$.
      \item $y^h_i-y^\ell_q>y^\ell_q$ and $\mu(h)>\frac{y^\ell_q}{2y^\ell_q-y^h_i+y^h_q}$.
    \end{enumerate}
  \end{enumerate}
\end{cor}

\end{appendix}
\end{document}